\documentclass[%
reprint,
nofootinbib,
 amsmath,amssymb,
 aps,
prd,
raggedbottom
]{revtex4-2}

\usepackage[color=blue, final]{changes}

\usepackage[]{putexModified}
\usepackage[]{hyperref}
\usepackage[percent]{overpic}
\usepackage{tikz}
\usetikzlibrary{decorations.pathreplacing}
\usepackage{subfigure}
\usetikzlibrary{calc,math} 
\usepackage{cancel}

\makeatletter
\DeclareRobustCommand*{\bfseries}{%
  \not@math@alphabet\bfseries\mathbf
  \fontseries\bfdefault\selectfont
  \boldmath
}
\makeatother

\input{glyphtounicode}
\begin{document}

\title{Toward Holography on Biregular Trees}

\author{Arkapal Mondal}%
  \email{phz228027@physics.iitd.ac.in}
\author{Sarthak Parikh}%
 \email{sarthak@physics.iitd.ac.in}
\author{Pulak Pradhan}%
  \email{pulakp133@gmail.com}
\author{Ritu Sengar}%
  \email{ritusengar2021@gmail.com}
\affiliation{%
  Department of Physics, Indian Institute of Technology Delhi, Hauz Khas, New Delhi 110016, India
}%
 
\date{\today}

\begin{abstract}
    We study scalar field theory on biregular trees, as a new model for discrete holography. Biregular trees are discrete symmetric spaces associated with the bulk isometry group ${\rm SU}(3)$ over the unramified quadratic extension of a nonarchimedean field. 
    The bulk-to-bulk and bulk-to-boundary propagators exhibit distinct features absent on the regular tree or continuum AdS spaces, arising from the semihomogeneous nature of the bulk space. 
    We compute the two- and three-point correlators of the putative boundary dual. 
    The three-point correlator exhibits a nontrivial ``tensor structure'' via dependence on the homogeneity degree of a unique bulk point specified in terms of boundary insertion points. 
    The computed OPE coefficients show dependence on zeta functions associated with the unramified quadratic extension of a nonarchimedean field. 
    This work initiates the formulation of holography on a family of discrete holographic spaces that exhibit features of both flat space and negatively curved space.
\end{abstract}

\maketitle


\section{Introduction}
\label{sec:INTRO}

The Anti-de Sitter/Conformal Field Theory (AdS/CFT) correspondence has led to remarkable insights into quantum gravity and strongly coupled field theories. 
In the last ten years, there has been growing interest in discrete formulations of holography, both from the point of view of tensor networks~\cite{Pastawski:2015qua} as well as effective field theories on tree lattices~\cite{Gubser:2016guj,Heydeman:2016ldy}.\footnote{For other related approaches to discrete holography, see, {\it e.g.} Refs.~\cite{Hung:2019zsk,Axenides:2019lea,Basteiro:2022zur}.} 
Discretisation offers both computational simplicity and a natural regularisation, while preserving key features of holography.
In fact, tree lattice discretisations naturally lead to a holographic direction in the bulk, making them especially amenable to a holographic correspondence.

In this work, we propose a large class of discrete holographic spaces arising from Bruhat--Tits buildings, which are simplicial complexes that correspond to discrete analogues of symmetric spaces~\cite{BruhatTits1972,serre1980trees,Abramenko_Brown_2008}. 
The simplest example for the bulk is the Bruhat--Tits tree associated with ${\rm PGL}(2,\mathbb{Q}_p)$, where $\mathbb{Q}_p$ is the field of $p$-adic numbers, which is a $(p+1)$-regular tree.
The regular tree comes with a natural notion of boundary with a $p$-adic ultrametric structure. 
This led to the formulation of $p$-adic AdS/CFT~\cite{Manin:2002hn,Gubser:2016guj,Heydeman:2016ldy}, where the boundary theory is a nonlocal $p$-adic conformal field theory.

In this work, we extend such holographic constructions to the setting of biregular trees.\footnote{Biregular trees have also recently appeared in the literature in constructing ``golden gates'' and efficient approximations of arbitrary quantum gates for fault-tolerant quantum computing~\cite{sarnak2015letter,Evra2022}.} 
These arise as Bruhat-Tits buildings associated with the unitary group over $p$-adic numbers, corresponding to subcomplexes of the Bruhat--Tits building for ${\rm PGL}(3,\mathbb{Q}_{p^2})$ over the unramified quadratic extension of $p$-adic numbers~\cite{Cartwright_Steger_1998}. 
Theories defined on these trees exhibit novel features absent in the regular tree case.

Our main technical results are:
\begin{itemize}
\item We derive explicit forms for bulk-to-bulk and bulk-to-boundary propagators on biregular trees, both in bulk and  boundary coordinates, showing how the semihomogeneous nature of the bulk introduces degree-dependent factors absent in the regular tree case. We also derive the mass-dimension relation and the associated BF bound.
\item We holographically compute two- and three-point correlators of the putative dual ${\rm SU}(3)$ ``conformal’’ theory on the boundary. The three-point correlator exhibits a nontrivial ``tensor structure'' that depends on the degree of the unique bulk vertex where geodesics from the three boundary insertion points intersect. We also furnish closed-form expressions for the three-point coefficient in terms of local zeta functions.
\item We establish two- and three-propagator identities in the bulk that express
 products of bulk-to-bulk propagators integrated over a common point in terms of unintegrated products of propagators. Such identities can help evaluate higher-point diagrams.  
\end{itemize}

The paper is organised as follows. In Sec.~\ref{sec:TREE}, we introduce the biregular tree and important properties of its boundary, and briefly summarise its algebraic origin from Bruhat-Tits theory, including the connection to flat space. 
In Sec.~\ref{sec:PROPAGATORS} we present the calculation of bulk-to-bulk and bulk-to-boundary propagators, as well as the two-point function and the behaviour of bulk fields near the boundary. 
In Sec.~\ref{sec:THREEPT}, we compute the three-point correlator from the bulk theory. 
In Sec.~\ref{sec:PROPIDS}, we derive two- and three-propagator bulk identities.
We conclude by discussing future directions in Sec.~\ref{sec:DISCUSS}. 
Readers not interested in or familiar with Bruhat--Tits theory and $p$-adic numbers may skip Sec.~\ref{sec:PADIC}, and still readily follow the main results in subsequent sections.

\section{Biregular trees}
\label{sec:TREE}

\begin{figure}[t]
\centering
\begin{tikzpicture}[scale=0.8,
  every node/.style={circle,inner sep=1.5pt},
  rnode/.style={fill=red},
  bnode/.style={fill=black}]

\def\R{3.2}
\draw (0,0) circle (\R);

\def\rA{.8}   
\def\rB{1.4}   
\def\rC{1.9}   
\def\rD{2.3}  

\node[bnode] (c) at (0,0) {};

\foreach \i/\ang in {1/0, 2/90, 3/180, 4/270}{
  \node[rnode] (d1\i) at (\ang:\rA) {};
  \draw (c) -- (d1\i);

  \foreach \j/\del in {1/-21, 2/21}{
     \pgfmathsetmacro{\angTwo}{\ang+\del}
     \node[bnode] (d2\i\j) at (\angTwo:\rB) {};
     \draw (d1\i) -- (d2\i\j);

     \foreach \k/\dthree in {1/-15, 2/0, 3/15}{
        \pgfmathsetmacro{\angThree}{\angTwo+\dthree}
        \node[rnode] (d3\i\j\k) at (\angThree:\rC) {};
        \draw (d2\i\j) -- (d3\i\j\k);

        \foreach \m/\dfour in {1/-3.0, 2/3.0}{
           \pgfmathsetmacro{\angFour}{\angThree+\dfour}
           \node[bnode] at (\angFour:\rD) {}; 
           \draw (d3\i\j\k) -- (\angFour:\rD);
        }
     }
  }
}
\def\rEllipsis{2.75} 
\node[inner sep=0pt] at (0:\rEllipsis) {$\cdots$};
\node[inner sep=0pt] at (180:\rEllipsis) {$\cdots$};
\node[inner sep=0pt,rotate=90] at (90:\rEllipsis) {$\cdots$};
\node[inner sep=0pt,rotate=90] at (270:\rEllipsis) {$\cdots$};

\end{tikzpicture}
\caption{A finite subset of $(2+1,3+1)$-biregular tree $T_{2,3}$, marking vertices of alternating degrees with different colours. The boundary at infinity of the tree is schematically represented as a circle for illustration; for more details, see Sec.~\ref{sec:BDY}.}
\label{fig:biregular-tree}
\end{figure}
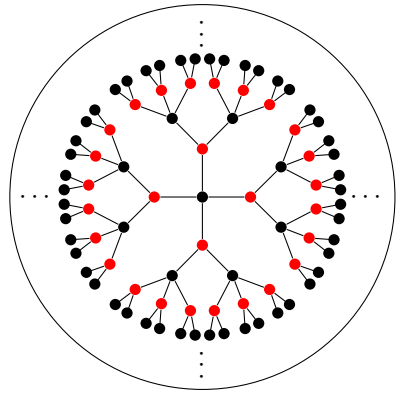

A tree is a cycle-free, connected, infinite graph.
A {\it regular} tree is a tree where each vertex has the same valency or degree, {\it i.e.} the number of edges connected to it.
It is a homogeneous, isotropic space. 
A {\it biregular} tree~\cite{Bouaziz-Kellil:1988,Talamanca_Nebbia_1991,gerardin2000,tarabusi2022} is a tree where every vertex has one of two possible degrees such that no two adjacent vertices have the same degree.
This tree is also sometimes referred to as the semihomogeneous or bihomogeneous tree, as the automorphisms of the tree no longer include odd translations along a geodesic due to the distinct degrees of neighbouring vertices.
Fig.~\ref{fig:biregular-tree} shows an example of a biregular tree where vertices have either degree $3$ or $4$. We call such a tree a $(2+1,3+1)$-biregular tree, and denote it $T_{2,3}$. 
In this paper, we will be working with the biregular tree $T_{q_+,q_-}$, for $q_\pm \geq 2$. When $q_+=q_-$, we recover the regular tree.

 In this paper, we fix a base vertex $v_0 \in T_{q_+,q_-}$. Then the {\it homogeneity degree} of vertex $a$, denoted $q_a$ is the number of outward neighbours of $a$ with respect to $v_0$. It can take one of two values, $q_\pm$.  In Fig.~\ref{fig:biregular-tree}, the black vertices have homogeneity degree $q_{\rm black}=3$ and red vertices have homogeneity degree $q_{\rm red}=2$.

\subsection{Boundary of biregular trees}
\label{sec:BDY}

(Bi)regular trees are examples of discrete spaces of non-positive curvature. For instance, using a graph-theoretic analogue of Ricci curvature~\cite{Gubser:2016htz}, it can be checked that they have constant negative curvature at all vertices.
In fact, they possess a more abstract notion of non-positive curvature, called Gromov-hyperbolicity, defined using Gromov products:
For points $a,b,c \in X$ in a metric space $(X,d)$, the {\it Gromov product} of $a$ and $b$ with respect to $c$ is defined to be
\eqn{GromProd}{
    (a,b)_c := \frac{1}{2}\left( d(a,c) + d(b,c) - d(a,b) \right).
}
The metric space is said to be $\delta$-hyperbolic if for all $x,y,z,w \in X$, $(x,z)_w \geq \min ((x,y)_w,(y,z)_w)-\delta$ for some $\delta>0$.
$X$ is called Gromov-hyperbolic if it is $\delta$-hyperbolic for some $\delta>0$.   
For instance, a tree is a Gromov-hyperbolic metric space where the metric $d$ is specified by the length of the shortest (non-backtracking) path between two vertices. 

One can associate a natural boundary with a Gromov-hyperbolic space $(X,d)$. 
A geodesic ray is an isometry, $r:[0,\infty) \to X$,  such that  $r([0,t])$ is the shortest path from the base point $r(0)=v_0$ to a point $r(t) \in X$ for all $t>0$. 
We say two geodesic rays $r_1, r_2$ are equivalent if they remain a bounded distance apart from each other for all $t>0$, {\it i.e.} $d(r_1(t),r_2(t)) < \infty$ for all $t$.
The {\it visual boundary} of $X$, denoted $\partial X$, is defined as the set of equivalence classes of geodesic rays starting at base point $v_0$. 
In fact, the visual boundary is independent of the choice of base point.

In the case of regular trees,\footnote{\added{The parameter $t$ in the definitions of the previous paragraph is discrete in this case.}} with $q_+=q_-=p$ a prime number, it is well known that points on the visual boundary $\partial T_{p,p}$ are in one-to-one correspondence with the set of $p$-adic numbers $\mathbb{Q}_p$ ~\cite{serre1980trees,gouvea2020padic} (more precisely, the projective line over the field of $p$-adic numbers, $\mathbb{P}^1(\mathbb{Q}_p)$) equipped with an ultrametric $p$-adic norm, where each $p$-adic number represents a unique path from the base vertex to the boundary at infinity (see {\it e.g.} Ref.~\cite{Gubser:2016guj}).
We will now describe the ultrametric nature of the boundary of biregular trees. 
In the next subsection, we will comment on algebraic constructions that give rise to biregular trees of a particular class of homogeneity degrees.

\begin{figure}[t]
\centering
\begin{tikzpicture}[scale=1.25]
  \draw (0,0) circle (1.5);
  
  \coordinate (x1) at (135:1.5);
  \coordinate (x2) at (0:1.5);
  \coordinate (x3) at (225:1.5);
  
  \coordinate (c) at (-0.5,0);
  
  \coordinate (b) at (0.6,0);
  \coordinate (a) at (0.6,0.6);
  
  \draw[black, very thick] (x1) -- (c) -- (b) -- (a);
  
  \draw[black, very thick] (x3) -- (c);
  
  \draw[black, very thick] (x2) -- (b);
  
  \fill (x1) circle (1.5pt) node[left] {\large $x_1$};
  \fill (x2) circle (1.5pt) node[right] {\large $x_2$};
  \fill (x3) circle (1.5pt) node[left] {\large $x_3$};
  
  \fill (c) circle (1.5pt) node[left] {\large $o$};
  \fill (b) circle (1.5pt) node[below] {\large $w$};
  \fill (a) circle (1.5pt) node[above] {\large $v_0$};
  \node at (45:1.9) { $\partial T_{q_+,q_-}$};

\end{tikzpicture}
  \caption{Geodesics on the biregular tree connecting boundary points $x_1, x_2, x_3$, intersecting at bulk point $o={\rm join}(x_1,x_2,x_3)$. The geodesic joining base vertex $v_0$ to the boundary points first intersects the other geodesics at $w$ as shown.}\label{fig:ultrametric}
 \end{figure}
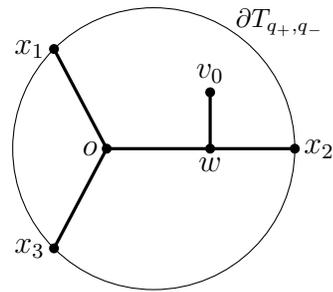

A key property of any tree is that there exists a unique point of intersection of geodesics connecting any three non-coincident points. Thus, setting $X=T_{q_+,q_-}$ and defining ${\rm join}(a,b,c)$ to be the unique point of intersection of geodesics between (bulk/boundary) points $a,b,c$, we have the obvious equality,
\eqn{GromProdExtend}{
    (a,b)_c = d(c,{\rm join}(a,b,c))  \,.
}
The right-hand-side allows us to extend the Gromov product to $a,b \in T_{q_+,q_-}\cup \partial T_{q_+,q_-}$.
We can then define a natural ultrametric on the boundary, known as the {\it visual metric}  
\eqn{VisualMetric}{
|x-y|_{\mathfrak{q}} := \mathfrak{q}^{-(x,y)_{v_0}} \,,
}
for any $x,y \in \partial T_{q_+,q_-}$ and fixed base vertex $v_0$, where we have introduced a new parameter\footnote{More generally, one can use the Gromov product to construct a natural visual metric on the visual boundary of any Gromov-hyperbolic space (see {\it e.g.} Ref.~\cite{Gesteau:2022hss}).}\textsuperscript{,}\footnote{\label{fn:PSHaar}The visual metric above is closely related to quasi-conformal measures one can define on the boundary, called Patterson-Sullivan measures, which behave as close to ``conformal'' as possible under the isometries of the space. 
 In fact, on the boundary of the regular tree $T_{p,p}$ with $q_+=q_-=p$ a prime, \eqref{VisualMetric} precisely corresponds to the measure of the smallest clopen ball on the boundary $\partial T_{p,p}=\mathbb{P}^1(\mathbb{Q}_p)$ containing both $x,y \in \partial T_{p,p}$ (see {\it e.g.} Ref.~\cite{Heydeman:2018qty}). 
 In the limit where the base point $v_0$ approaches the boundary, the Patterson-Sullivan measure reduces to the familiar Haar measure on $p$-adic numbers $\mathbb{Q}_p$~\cite{Heydeman:2018qty}.}
 \eqn{frakqDef}{ 
 \mathfrak{q} := \sqrt{q_+ q_-}\,.
 }
An {\it ultrametric} space satisfies a stronger form of triangle inequality,
\eqn{Ultrametric}{ 
|x_{ij}| \leq \max \{ |x_{jk}|, |x_{ki}| \}
}
for all $x_i, x_j, x_k$, where $x_{ij} := x_i-x_j$.
 We now show that the visual boundary of the biregular tree, $(\partial T_{q_+,q_-},~|~\cdot~|_\mathfrak{q})$, endowed with the visual metric defined above, is an ultrametric space. Without loss of generality, assume the geodesic configuration shown in Fig.~\ref{fig:ultrametric} for boundary points $x_1,x_2,x_3$. Then, it follows,
 \eqn{}{
 |x_{12}|_\mathfrak{q}  = |x_{23}|_\mathfrak{q} = \mathfrak{q}^{-d(v_0,w)} \geq |x_{31}|_\mathfrak{q} = \mathfrak{q}^{-d(v_0,w)-d(w,o)},
 }
 from which the ultrametricity property~\eqref{Ultrametric} is easily verified for all permutations of $i,j,k$.
 
\subsection{Algebraic origins}
\label{sec:PADIC}

In this section, we collect several relevant facts from the literature about the algebraic origins of biregular trees.
We omit a comprehensive and mathematically rigorous exposition for brevity, opting instead to provide references.
This section can be skipped on a first reading by readers seeking to go straight to the main results of our work.

There are profound similarities between Riemann symmetric spaces and {\it Euclidean buildings}; see {\it e.g.} Refs.~\cite{stadler2024cat0I,stadler2024cat0II}.
Euclidean buildings, also called {\it Bruhat--Tits buildings}, are combinatorial objects that assign a geometric meaning to various algebraic groups~\cite{BruhatTits1972,Abramenko_Brown_2008,Ji_2006}. 
For instance, the Bruhat--Tits building associated with the group ${\rm PGL}(2,\mathbb{Q}_p)$, also known as the Bruhat--Tits tree, is the $(p+1)$-regular tree~\cite{serre1980trees}, where $\mathbb{Q}_p$ is the field of $p$-adic numbers.\footnote{\label{fn:powers}One can also get regular trees of degree $p^n+1$, coming from various field extensions of $\mathbb{Q}_p$~\cite{Manin1976,Gubser:2016guj,Ghoshal:2006zh}.}
The group ${\rm PGL}(2,\mathbb{Q}_p)$ is the isometry group of the tree that acts on its vertices transitively. 
The action of the isometry group is most intuitive when the vertices are viewed as equivalence (homothety) classes of lattices in $\mathbb{Q}_p \times \mathbb{Q}_p$, equipped with a notion of incidence relation. 
The isometry group acts on the equivalence classes of lattices by matrix multiplication~\cite{Brekke:1993gf}.
These isometries include translations along a geodesic, rotations about a vertex, and inversions about an edge. 
A maximal compact subgroup stabilises a vertex; thus, the set of vertices of the tree can also be identified with the quotient of the isometry group by a maximal compact subgroup.
This leads to an interpretation of the tree as a symmetric space, over the nonarchimedean field of $p$-adic numbers.
The visual boundary of the tree is given by the projective line $\mathbb{P}^1(\mathbb{Q}_p)$. As discussed in the previous subsection, it has an ultrametric (nonarchimedean) nature and a totally disconnected topology.
The isometry group induces fractional linear transformations on the boundary.

The discussion above admits a generalisation to higher rank groups $G={\rm PGL}(n,\mathbb{Q}_p)$ for all $n \geq 3$. The Bruhat--Tits building associated with the rank $n$ group $G$ is a simply connected simplicial complex of dimension $n-1$.
It has maximal subcomplexes, called {\it apartments}, which are tessellations of $(n-1)$-dimensional Euclidean space, $\mathbb{R}^{n-1}$.
Thus, ``slices'' or sections of these buildings are flat spaces, while simultaneously the volume of a ball grows exponentially, indicating hyperbolicity.
The vertices of the simplicial complex are identified with the homothety classes of lattices in $\mathbb{Q}_p^{n}$, and the isometry group $G$ acts on them transitively via matrix multiplication.
The Bruhat--Tits building is a symmetric space with the vertex set isomorphic to the quotient space $G/K$, where $K$ is a maximal compact subgroup of $G$ that fixes a vertex.
The visual boundary of the building is a maximal flag complex~\cite{king2012k} which is a {\it Spherical building} (also called Tits building)~\cite{Borel_Ji_2006, Ji_2006,Ji2009,Abramenko_Brown_2008,caprace2015}, whose apartments are tessellations of $(n-1)$-sphere.
This boundary has a totally disconnected topology and an ultrametric nature.

For the present work, the relevant group is ${\rm PGL}(3,\mathbb{Q}_{p})$. Its building is a 2-dimensional simplicial complex whose apartments are triangulations of $\mathbb{R}^2$. Its edges have degree $p+1$, ({\it i.e.} $p+1$ triangles are incident at each edge), and its vertices have degree $2(p^2+p+1)$.
If $\mathbb{Q}_{p^2}$ denotes the unique unramified quadratic extension of $p$-adic numbers, then the building of ${\rm PGL}(3,\mathbb{Q}_{p^2})$ is a 2-dimensional simplicial complex, whose edges have degree $p^2+1$ and whose vertices have degree $2(p^4+p^2+1)$~\cite{Dalal:2025lzk,Cartwright_Steger_1998}.
It turns out, the $(p^3+1,p+1)$-biregular tree is a subcomplex of the Bruhat--Tits building associated with ${\rm PGL}(3,\mathbb{Q}_{p^2})$, as we now describe~\cite{Cartwright_Steger_1998,gerardin2000,Evra2022,Dalal:2025lzk}.

Let ${\rm PU}(3,\mathbb{Q}_p,h)$ denote the group of $3 \times 3$ unitary matrices modulo its center with entries in $\mathbb{Q}_{p^2}$, that preserves a hermitian form $h$ on $\mathbb{Q}_{p^2}^3$.
${\rm PU}(3,\mathbb{Q}_{p},h)$ is a subgroup of ${\rm PGL}(3,\mathbb{Q}_{p^2})$, and the Bruhat--Tits building associated with ${\rm PU}(3,\mathbb{Q}_{p},h)$ is a subcomplex. (In fact, the buildings associated with the isometry groups PU, U, and SU are the same~\cite{Cartwright_Steger_1998,Evra2022}.) 
In the lattice description of the building for ${\rm PGL}(3,\mathbb{Q}_{p^2})$ where vertices are identified with equivalence classes of lattices, there exists an adjacency-preserving and face-preserving involution that fixes one of the vertices of a triangular face and interchanges the other two.
Thus, we can define two sets of vertices: (i) the set of vertices fixed by the involution, and (ii) the set of two-tuples of vertices that get interchanged under the involution. 
Elements from these sets are called adjacent if together they form a face in the parent building.
These sets, along with the adjacency relation, form a graph which is precisely a $(p^3+1,p+1)$-biregular tree.
The vertices fixed by the involution have $p^3+1$ neighbours (and are called { hyperspecial}) while the two-tuple of vertices that get interchanged have $p+1$ neighbours (and are called { special}).
The isometry group ${\rm PU}(3,\mathbb{Q}_{p},h)$ acts compatibly on the vertices of the biregular tree via matrix multiplication.
(For an explicit and rigorous account of this Bruhat--Tits building, see Ref.~\cite{Cartwright_Steger_1998}.\footnote{ See also Ref.~\cite{Dalal:2025lzk} for a construction of $(p^2+1,p+1)$-biregular trees as ${\rm PU}(4)$ subcomplexes of buildings associated with ${\rm PGL}(4)$, and Ref.~\cite{cappellini2020} for a construction of the $(5,3)$-biregular tree as the Bruhat--Tits building for a subgroup of ${\rm PGL}{}(5,\mathbb{Q}_2)$ that preserves a quadratic form of signature $(4,1)$.}) 
The visual boundary of the biregular tree is a hermitian flag complex, a Spherical building of totally disconnected topology and ultrametric nature~\cite{caprace2015}. 

Note that the automorphism group of the biregular tree is not vertex transitive and has two orbits, corresponding to the two sets of vertices above, of different homogeneity degrees. 
Consequently, compared to the regular tree,  other than rotations, only even translations are permitted as automorphisms~\cite{Talamanca_Nebbia_1991}.

In this paper, we’ll work with generic biregular trees, not just those known to emerge as the Bruhat--Tits building of nonarchimedean unitary groups. Thus, our results will apply to all biregular trees.

\section{Propagators}
\label{sec:PROPAGATORS}

We start with the simplest bulk theory on the biregular tree $T_{q_+,q_-}$, described by a quadratic lattice action for a massive scalar field living on the vertices of the tree,
\eqn{Squadratic}{
S[\phi] =  \sum_{\langle ab \rangle} \frac{1}{ 2} (\phi_a - \phi_b)^2 + \sum_{a \in T_{q_+,q_-}} \!\!\!\!\left( \frac{1}{2} m_{\Delta}^2 \phi_a^2  - J_a \phi_a \right).
}
The action is identical to the one considered in the simplest versions of $p$-adic AdS/CFT on $(p+1)$-regular trees~\cite{Gubser:2016guj}, with the exception that the space over which the bulk sums are taken has changed to a biregular tree.
In~\eqref{Squadratic}, the first term, summed over all nearest neighbour vertices $a,b$ (or equivalently, all edges), is the standard graph discretisation of the kinetic derivative term. 
The subscript in $m^2_{\Delta}$ is introduced for future convenience, to be identified with the dual scaling dimension.
The Euler-Lagrange equation of motion obeyed by the scalar field $\phi$ is
\eqn{EOM}{
(\square_a + m_{\Delta}^2) \phi_a = J_a \,,
}
where $\square_a$ is the graph vertex Laplacian acting on the $a$ vertex, defined as
\eqn{Laplacian}{
\square_a \phi_a = \sum_{c \sim a} (\phi_a-\phi_c)\,,
}
where the sum over $c \sim a$ denotes a sum over all nearest-neighbours $c$ of $a$.
Depending on the homogeneity degree of $a$, this evaluates to
\eqn{Laplace}{
\square_a \phi_a = (q_a+1)\phi_a - \sum_{c \sim a} \phi_c \,.
}
The equation of motion can be solved in terms of the source and the Green's function $G_{\Delta}$,
\eqn{}{
\phi_a = \sum_{b \in T_{q_+,q_-}} G_{\Delta}(a,b) J_b\,,
}
which  satisfies\footnote{The Kronecker delta $\delta_{ab}$ equals unity for coincident bulk points $a,b$ and vanishes otherwise.}
\eqn{GreensEOM}{
(\square_a + m_\Delta^2) G_{\Delta}(a,b) = \delta_{ab}\,,
}
with appropriate boundary conditions.

\subsection{Bulk-to-bulk propagator}
\label{sec:GREENS}

The Green's function between bulk points $a$ and $b$ is a bulk-to-bulk propagator that plays a central role in perturbative treatments of interacting theories via bulk Feynman diagrams. 
In this section, we will explicitly derive its form on the biregular tree.
As we demonstrate, the propagator exhibits distinct features absent on the regular tree or continuum AdS spaces, arising from the semihomogeneous nature of the bulk space.

Fix a vertex $b \in T_{q_+,q_-}$. We seek spherically symmetric solutions to~\eqref{GreensEOM}, that is, solutions of the form, $G_{\Delta}(a,b) = g(d)$ which depend only on the geodesic distance $d:=d(a,b)$.
For bulk vertices $a \neq b$, {\it i.e.} $d\geq 1$, substituting this form in~\eqref{GreensEOM}, we get 
\eqn{aNotb}{
(q_a+1+m_{\Delta}^2) g(d) - q_a g(d+1) - g(d-1) = 0\,,
}
where we used the fact that $q_a$ nearest-neighbours of vertex $a$ are farther away from $b$ than $a$ by one unit of distance, while exactly one nearest-neighbour of $a$ is one step closer to $b$.
Importantly, the coefficients of the second-order difference equation~\eqref{aNotb} are $d$-dependent since $q_a$ alternates between $q_+$ and $q_-$ as the parity of $d$ alternates.
Indeed,~\eqref{aNotb} comprises a pair of two difference equations with constant coefficients,
\eqn{aNotbAgain}{
(q_b+1+m_{\Delta}^2) g(2k+2) - q_b g(2k+3) - g(2k+1) &= 0  \cr 
(\tilde{q}_b+1+m_{\Delta}^2) g(2k+1) - \tilde{q}_b g(2k+2) - g(2k) &= 0 
}
for $k=0,1,2,\ldots$, where $\tilde{q}_b$ is the homogeneity degree of any nearest-neighbour of $b$.\footnote{We denote by $\tilde{b}$  any nearest-neighbor of $b$ and define $\tilde{q}_b$ to be the opposite homogeneity degree to $q_b$, so $ q_{\tilde{b}} = \tilde{q}_b$. Naturally, if $q_b=q_+$, then $\tilde{q}_b=q_-$.
\added{Moreover,
\eqn{qaFormula}{
q_a = \frac{1+(-1)^{d(a,b)}}{2} q_b + \frac{1-(-1)^{d(a,b)}}{2} \tilde{q}_b\,.
}}}
This can be understood as a consequence of the fact that the adjacency operator (Laplace minus the diagonal operator in~\eqref{Laplace}) is not transitive on the set of vertices.

The adjacency operator mixes vertices of different homogeneity degrees belonging to separate orbits.
However, its square preserves homogeneity; thus, it acts on the two orbits of the automorphism group separately, leading to step-2 transitions in the associated difference equations.
One can obtain these equations by substituting for $g(2k+1)$ and $g(2k+3)$ using the second difference equation in~\eqref{aNotbAgain}. Then, the first difference equation becomes 
\eqn{}{
& \left[(q_b+1+m_{\Delta}^2)(\tilde{q}_b+1+m_{\Delta}^2) - q_b-\tilde{q}_b\right] g(2k+2) \cr 
&- q_b \tilde{q}_b g(2k+4)
- g(2k) = 0 \,.
}
This equation is symmetric in $q_b$ and $\tilde{q}_b$, so can be readily rewritten as
\eqn{gEven}{
& \left[(q_++1+m_{\Delta}^2)({q}_-+1+m_{\Delta}^2) - q_+-{q}_-\right] g(2k+2) \cr 
&- q_+q_- g(2k+4)
- g(2k) = 0 \,.
}
Likewise, one can isolate an identical difference equation for the other orbit, corresponding to the function $g$ with solely odd integral arguments,
\eqn{gOdd}{
& \left[(q_++1+m_{\Delta}^2)({q}_-+1+m_{\Delta}^2) - q_+-{q}_-\right] g(2k+3) \cr 
&- q_+q_- g(2k+5)
- g(2k+1) = 0 \,.
}
We recover the same step-2 difference equation in each orbit. 
Combining, we obtain the difference equation,
\eqn{gDiffEq}{
& \left[(q_++1+m_{\Delta}^2)({q}_-+1+m_{\Delta}^2) - q_+-{q}_-\right] g(d+2) \cr 
&- q_+q_- g(d+4)
- g(d) = 0 \,,
}
for all $d\geq 0$. A solution is given by
\eqn{gSol}{
g(d) = \left(C_1 + (-1)^d C_2\right) \mathfrak{q}^{-\Delta d}\,,
}
where $\mathfrak{q}$ was defined in~\eqref{frakqDef} and the constants $C_1,C_2$ will be determined shortly. 
 We have implicitly defined the scaling dimension $\Delta$ in terms of the mass parameter, using
 \eqn{mass-dim-pm}{
(q_+\!+\! 1\!+m_\Delta^2)({q}_-\!+\! 1\!+m_\Delta^2)  =  ({q}_+^{\Delta} + {q}_-^{1-\Delta} )  ({q}_-^{\Delta} + q_+^{1-\Delta} ) \,.
}
For $m_{\Delta}^2>0$, there are two real solutions for the scaling dimension, namely $\Delta_+>1$ and $\Delta_-<0$ such that $\Delta_++\Delta_-=1$. Thus, seeking solutions that vanish at infinity, we discard the $\Delta_-$ solution.
In fact, small negative values of $m_{\Delta}^2$ are allowed as well, within the range $0>m_{\Delta}^2 \geq m_{\rm BF}^2$, such that both $\Delta_{\pm}$ are real and positive with $\Delta_++\Delta_-=1$. The Breitenlohner-Freedman (BF) bound is achieved at $\Delta_\pm=1/2$, corresponding to
\eqn{BFbound}{
m_{\rm BF}^2 &:= -\frac{q_++q_-+2}{2} \cr 
&+ \frac{\sqrt{q_+}+\sqrt{q_-}}{2}  \sqrt{(\sqrt{q_+}-\sqrt{q_-})^2+4}\,.
}
In this paper, we will set $\Delta=\Delta_+ >1/2$ and refrain from discussing alternative quantisation. 

To determine the constants, we substitute~\eqref{gSol} into~\eqref{aNotbAgain} to get an equation relating $C_1$ and $C_2$.\footnote{Naively, one might think~\eqref{aNotbAgain} furnishes two equations, but they are mathematically equivalent due to the mass-dimension relation~\eqref{mass-dim-pm}.}
We get the second relation by imposing the boundary condition $a=b$ in~\eqref{GreensEOM}. Solving for $C_1, C_2$, we obtain the unique solution
\eqn{Gcases}{
& G_{\Delta}(a,b)/N_{\Delta}  \cr &= \begin{cases}
     \sqrt{\displaystyle{\frac{(\tilde{q}_b+1+m_{\Delta}^2)(q_b+\mathfrak{q}^{2\Delta})}{(q_b+1+m_{\Delta}^2)(\tilde{q}_b+\mathfrak{q}^{2\Delta})}}}  \mathfrak{q}^{-\Delta d(a,b)}  & d(a,b) {\rm \ even} \cr 
       \mathfrak{q}^{-\Delta d(a,b)} & d(a,b) {\rm \ odd} 
           \end{cases}
}
where the normalisation constant is defined to be
\eqn{NDef}{
N_{\Delta} := \frac{1}{\mathfrak{q}^\Delta-\mathfrak{q}^{-\Delta}}\,,
}
and we employed an alternate, useful form of the biregular tree mass-dimension relation,
\eqn{mass-dim}{
  \mathfrak{q}^{2\Delta}\, (q_b\!+\! 1\!+m_\Delta^2)(\tilde{q}_b\!+\! 1\!+m_\Delta^2)  =   (q_b+ \mathfrak{q}^{2\Delta}) (\tilde{q}_b+\mathfrak{q}^{2\Delta}),
}
for any vertex $b$.
It will be convenient to rewrite the bulk-to-bulk propagator without separating the cases, in the final form 
\eqn{G}{
G_{\Delta}(a,b) = N_{\Delta} \frac{\psi_\Delta(\tilde{q}_a) }{ \psi_\Delta(q_b) }  \, \mathfrak{q}^{-\Delta\, d(a,b)} \,,
}
where we have defined
\eqn{psiDef}{  
\psi_\Delta(q_a) := \sqrt{ \frac{q_a+1+m_\Delta^2}{q_a+\mathfrak{q}^{2\Delta}} }
 \,.
}
The mass-dimension relation~\eqref{mass-dim} can be compactly expressed as
\eqn{mass-dim-psi}{
\psi_\Delta(q_a) \, \psi_\Delta(\tilde{q}_a)= \frac{1}{\mathfrak{q}^\Delta} \,.
}
This relation proves quite handy in calculations. Using this, we can also readily verify that $G_{\Delta}(a,b)$ is symmetric in $a$ and $b$, since we may write
\eqn{GAgain}{
G_{\Delta}(a,b) = \frac{N_{\Delta} }{ \psi_\Delta(q_a)\psi_\Delta(q_b) }  \, \mathfrak{q}^{-\Delta\,(d(a,b)+1)} \,.
}

Some comments are in order. \added{Naively, the assumed rotational invariance may not seem manifest in~\eqref{GAgain}, but it is indeed so, due to~\eqref{qaFormula}.} \replaced{Alternately,}{The} rotational invariance of the bulk-to-bulk propagator is manifest in~\eqref{Gcases}.
Furthermore, that the automorphism group admits only even translations is reflected in the fact that the bulk-to-bulk propagator depends on the homogeneity degree of one of the end-points and the parity of the distance between the two bulk points.

It is instructive to compare the bulk-to-bulk propagator on regular (homogeneous) and biregular (semihomogeneous) trees. On a $(\mathfrak{q}+1)$-regular tree,\footnote{On the regular tree, we assume $\mathfrak{q}$ is an integer $\geq 2$.} the propagator takes the form~\cite{Gubser:2016guj,Heydeman:2016ldy}
\eqn{Gregular}{
G_{\Delta}^{\rm (regular)}(a,b) = N_{\Delta} \mathfrak{q}^{-\Delta\, d(a,b)}\,,
}
where  the mass-dimension relation is given by
\eqn{mass-dim-regular}{
m_\Delta^{2} = -1 - \mathfrak{q} + \mathfrak{q}^\Delta + \mathfrak{q}^{1-\Delta}\,.
}
This result follows immediately from the biregular tree expressions~\eqref{G} and~\eqref{mass-dim-pm}
upon taking the regular tree limit, $q_+=q_-=\mathfrak{q}$.\footnote{The astute reader might have noticed that in bulk of the treatments of $p$-adic AdS/CFT~\cite{Gubser:2016guj,Gubser:2016htz,Gubser:2017tsi,Gubser:2017vgc,Jepsen:2018dqp,Jepsen:2018ajn,Jepsen:2019svc}, when $\mathfrak{q}$  can be expressed as a power of a prime as alluded to in footnote~\ref{fn:powers}, the degree of the extension of the $p$-adic numbers appears as an explicit parameter. 
In contrast, we have made no such distinction here. 
This is due to a different choice of convention, $\Delta_{\rm here} = \Delta_{\rm there}/n$, where $n$ is the degree of unramified field extension of the $p$-adic field. Our convention matches the one described in Ref.~\cite[footnote 11]{Gubser:2017qed}.}

For future convenience, we also define an unnormalized propagator,
\eqn{Ghat}{
\hat{G}_{\Delta}(a,b) :=  \frac{\psi_\Delta(\tilde{q}_a) }{ \psi_\Delta(q_b) }  \, \mathfrak{q}^{-\Delta\, d(a,b)} \,.
}

\subsection{Bulk-to-boundary propagator}
\label{sec:BULKBDYPROP}

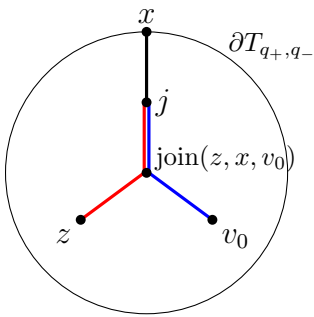
\begin{figure}[t]
\centering
\begin{tikzpicture}[scale=1.25]
  \draw (0,0) circle (1.5);
  
  \coordinate (x) at (90:1.5); 
  \coordinate (z) at (-0.7,-0.5); 
  \coordinate (v0) at (0.7,-0.5); 
  \coordinate (c) at (0,0);    
  \coordinate (j) at (0,0.75);  
  
  \coordinate (c1) at (-0.025,0);    
  \coordinate (c2) at (0.025,0);     
  \coordinate (j1) at (-0.025,0.75); 
  \coordinate (j2) at (0.025,0.75);  

  \draw[red, very thick] (z) -- (c1) -- (j1);

  \draw[blue, very thick] (v0) -- (c2) -- (j2);
  
  \draw[black, very thick] (j) -- (x);
  
  \fill (x) circle (1.5pt) node[above] {\large $x$};
  \fill (z) circle (1.5pt) node[below left] {\large $z$};
  \fill (v0) circle (1.5pt) node[below right] {\large $v_0$};
  \fill (c) circle (1.5pt) node[above right, yshift=-3pt] { $\mathrm{join}(z,x,v_0)$};
  \fill (j) circle (1.5pt) node[right] {\large $j$};
\node at (45:1.9) { $\partial T_{q_+,q_-}$};
\end{tikzpicture}
\caption{The horospherical index of $z$ with respect to $v_0$ and $x$, $\langle z,x \rangle_{v_0} = d(v_0,{\rm join}(z,x,v_0)) - d(z,{\rm join}(z,x,v_0))$.}
\label{fig:join}
\end{figure}

To calculate correlators in the putative boundary theory dual to the bulk model, we also need bulk-to-boundary propagators. 
The bulk-to-boundary propagator between bulk vertex $z \in T_{q_+,q_-}$ and boundary point $x \in \partial T_{q_+,q_-}$ can be obtained from the bulk-to-bulk propagator in the limit one of the bulk points goes to the boundary. It satisfies the equation of motion
\eqn{Keom}{
(\square_z + m^2_\Delta)K_\Delta(z,x) = 0 \,.
}
For a fixed base vertex $v_0$, the solutions to~\eqref{Keom} are proportional to 
\eqn{Khat}{
\hat{K}_\Delta(z,x) := \frac{\psi_\Delta(\tilde{q}_z)}{\psi_\Delta(\tilde{q}_{v_0})}\, \mathfrak{q}^{\Delta \langle z,x \rangle_{v_0}}\,,
}
where the {\it horospherical index} of $z$ with respect to $v_0$ and $x$ is defined to be~\cite{Cartier1972,Zabrodin1989}
\eqn{HorInd}{
\langle z,x \rangle_{v_0} := d(v_0,j) - d(z,j)\,,
}
where  $z$ is a bulk vertex, $x \in  T_{q_+,q_-} \cup \partial T_{q_+,q_-}$ can in general be either a bulk vertex or a boundary point, and the vertex $j \in [z:x) \cap [v_0:x)$.\footnote{The unique geodesic joining any two points $a$ and $b$ where $a, b \in T_{q_+,q_-} \cup \partial T_{q_+,q_-}$ is represented by a sequence of vertices along the shortest (non-backtracking) path connecting $a$ with $b$, denoted $(a:b)$ or $[a:b)$ depending on whether we include the end point in the path.}  Clearly, the horospherical index is independent of the choice of $j$. Recalling that ${\rm join}(a,b,c)$ refers to the unique point of intersection of geodesics joining bulk/boundary points $a,b,c$, we note that we can freely replace $j$ with ${\rm join}(z,x,v_0)$ in~\eqref{HorInd} since ${\rm join}(z,x,v_0)$ is the unique vertex $\in [z:x) \cap [v_0:x)$ that is closest to the vertices $z$ and $v_0$ (see Fig.~\ref{fig:join}). 

It is straightforward to show~\eqref{Khat} solves~\eqref{Keom}. Plugging in the solution on the left-hand-side, we obtain
\eqn{}{
& \left(q_z+1+m_{\Delta}^2\right) \frac{\psi_{\Delta}(\tilde{q}_{z})}{\psi_{\Delta}(\tilde{q}_{v_0})}  \mathfrak{q}^{\Delta \langle z,x \rangle_{v_0}} \cr 
&- \left(q_z \mathfrak{q}^{-\Delta} + \mathfrak{q}^{\Delta }  \right) \frac{\psi_{\Delta}({q}_{z})}{\psi_{\Delta}(\tilde{q}_{v_0})}  \mathfrak{q}^{\Delta \langle z,x \rangle_{v_0}}\,,
}
where we used the fact that the horospherical index of a nearest-neighbour of $z$ with respect to $v_0$ and $x$ will be shifted up or down one unit, depending on whether the nearest-neighbour is closer to $v_0$ or not, respectively. 
Using~\eqref{psiDef}-\eqref{mass-dim-psi}, it follows immediately that the above combination vanishes identically.

In the regular tree limit $q_+=q_-$,~\eqref{Khat} manifestly reduces to the bulk-to-boundary propagator for the regular tree derived in Ref.~\cite{Heydeman:2016ldy}, since the overall ratio of $\psi_\Delta$ factors in~\eqref{Khat} goes to unity. 
However, the choice of normalisation is different. The propagator~\eqref{Khat} is normalized such that $\hat{K}_\Delta(v_0,x) = 1$.\footnote{Moreover, it is related to the Green's function~\eqref{G} via~\cite{tarabusi2022}
\eqn{}{
\hat{K}_\Delta(z,x) = \frac{G_\Delta(z,{\rm join}(z,x,v_0))}{G_\Delta(v_0,{\rm join}(z,x,v_0))} \,.
}
}
We will comment further on the choice of normalisation in Sec.~\ref{sec:TWOPT}.

\subsection{Cross-ratios and propagators}
\label{sec:CROSSRATIO}

Although the presented form of propagators facilitates efficient computations in evaluating bulk diagrams, it is also helpful to work with an alternate form involving boundary cross-ratios explicitly. 
In this section, we present the alternate representations along the lines of similar representations for propagators on the regular tree~\cite{Gubser:2016guj}.

\begin{figure}[t]
 \subfigure[]{
\begin{tikzpicture}[scale=1]
  \draw (0,0) circle (1.5);
  
  \coordinate (x1) at (150:1.5);
  \coordinate (x2) at (210:1.5);
  \coordinate (x3) at (30:1.5);
  \coordinate (x4) at (330:1.5);
  
  \coordinate (a) at (-0.7,0.0);
  \coordinate (b) at (0.7,0.0);
  
  \draw[black, very thick] (x1) -- (a) -- (x2);
  \draw[black, very thick] (x3) -- (b) -- (x4);
  \draw[black, very thick] (a) -- (b);
  
  \fill (x1) circle (1.5pt) node[left] {\large$x_1$};
  \fill (x2) circle (1.5pt) node[left] {\large$x_2$};
  \fill (x3) circle (1.5pt) node[right] {\large$x_3$};
  \fill (x4) circle (1.5pt) node[right] {\large$x_4$};
  
  \fill (a) circle (1.5pt) node[above right] {\large$a$};
  \fill (b) circle (1.5pt) node[ above left] {\large$b$};

 \node at (45:2) { $\partial T_{q_+,q_-}$};
\end{tikzpicture}
\label{fig:cross-ratio}
}
 \subfigure[]{
 \begin{tikzpicture}[scale=1]
  \draw (0,0) circle (1.5);
  
  \coordinate (x) at (90:1.5);
  \coordinate (v) at (330:1.5);
  \coordinate (u) at (210:1.5);
  \coordinate (j) at (0,0.5);
  \coordinate (v0) at (-0.5,0.5);
  
  \coordinate (z) at (0,0);
  
  \draw[black, very thick] (x) -- (z) -- (v);
  \draw[black, very thick] (z) -- (u);
  \draw[black, very thick] (v0) -- (j);
  
  \fill (x) circle (1.5pt) node[above] {\large$x$};
  \fill (v) circle (1.5pt) node[right] {\large$v$};
  \fill (u) circle (1.5pt) node[left] {\large$u$};
  
  \fill (z) circle (1.5pt) node[below] {\large$z$};
  \fill (j) circle (1.5pt) node[right] {${\rm join}(z,x,v_0)$};
  \fill (v0) circle (1.5pt) node[left] {\large$v_0$};
  
 \node at (45:2) { $\partial T_{q_+,q_-}$};
 \label{fig:Kxuv}
\end{tikzpicture}
 }
\caption{(a): A configuration of four boundary points and their bulk interpretation. (b): A configuration of three boundary points.}

\end{figure}
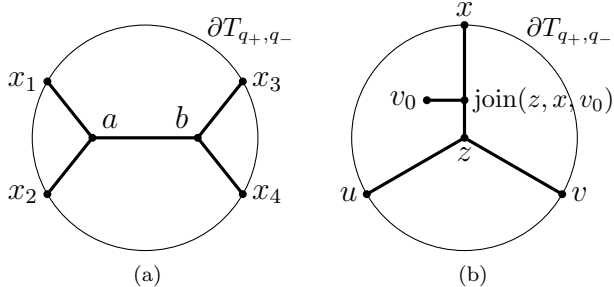

Given bulk points $a,b \in T_{q_+,q_-}$, extend two geodesic rays originating from $a$ corresponding to non-coincident boundary points $x_1$ and $x_2$, such that  $a = {\rm join}(x_1,x_2,b)$. 
Likewise, extend two geodesic rays originating at $b$ which reach the boundary at non-coincident points $x_3$ and $x_4$, with $b={\rm join}(x_3,x_4,a)$. 
See Fig.~\ref{fig:cross-ratio} for a visual representation of this configuration. 
Note that the choice of boundary points $x_1,x_2,x_3,x_4$ is highly non-unique, but any such choice satisfies the following relation, 
\eqn{uCrossratio}{
\mathfrak{q}^{-d(a,b)} = \frac{|x_{12}|_\mathfrak{q} |x_{34}|_\mathfrak{q}}{|x_{13}|_\mathfrak{q} |x_{24}|_\mathfrak{q}} \,,
}
where we expressed the right-hand-side in terms of the visual metric~\eqref{VisualMetric}. The formula establishes a bulk/boundary correspondence between the geodesic distance $d(a,b)$ and a boundary cross-ratio.
The proof is straightforward. Assuming the base vertex $v_0$ to be coming off of any of the five geodesic legs shown, one can express the right-hand-side in terms of graph distances via the Gromov product representation of the visual metric. In all cases, various factors cancel out, leaving behind the quantity on the left-hand-side.
The following equality also holds
\eqn{}{
\frac{|x_{14}|_\mathfrak{q} |x_{23}|_\mathfrak{q}}{|x_{13}|_\mathfrak{q} |x_{24}|_\mathfrak{q}} = 1\,,
}
corresponding to the fact that only one independent (nontrivial) cross-ratio of four points exists in an ultrametric space.

Using~\eqref{uCrossratio}, one can write the unnormalized bulk-to-bulk propagator entirely in terms of boundary data as
\eqn{Gbdy}{
\hat{G}_{\Delta}(a,b) = \frac{\psi_\Delta(\tilde{q}_{{\rm join}(x_1,x_2,x_3)}) }{ \psi_\Delta(q_{{\rm join}(x_3,x_4,x_1)}) }  \left(\frac{|x_{12}|_\mathfrak{q} |x_{34}|_\mathfrak{q}}{|x_{13}|_\mathfrak{q} |x_{24}|_\mathfrak{q}}\right)^\Delta,
}
for the configuration of points depicted in Fig.~\ref{fig:cross-ratio}.

Likewise, one can express the bulk-to-boundary propagator in terms of boundary data, generalising the regular tree result of Ref.~\cite{Gubser:2016guj}.
Fix a boundary point $x$ and a bulk vertex $z$. 
Without loss of generality, the base vertex $v_0$ will be located as schematically depicted in Fig.~\ref{fig:Kxuv}.\footnote{The case where ${\rm join}(z,x,v_0) = z$ is partially subsumed in the general configuration shown in Fig.~\ref{fig:Kxuv}, except when $v_0 \in (z:u)$ or $(z:v)$. In such a case, one can check that~\eqref{Kbdy} holds just as well.} 
Like before, extend two geodesic rays emanating from $z$ to reach non-coincident boundary points $u,v$ with the condition that $z={\rm join}(x,u,v)$. From Fig.~\ref{fig:Kxuv}), we observe,
\eqn{}{
|u-v|_\mathfrak{q} &= \mathfrak{q}^{-d(v_0,{\rm join}(z,x,v_0))-d({\rm join}(z,x,v_0),z)} \cr 
 |x-u|_\mathfrak{q} &= \mathfrak{q}^{-d(v_0,{\rm join}(z,x,v_0))} 
 \cr 
 |x-v|_\mathfrak{q} &= \mathfrak{q}^{-d(v_0,{\rm join}(z,x,v_0))} \,,
}
from which it follows
\eqn{}{
\frac{|u-v|_\mathfrak{q}}{|x-u|_\mathfrak{q} |x-v|_\mathfrak{q}} = \mathfrak{q}^{\langle z,x\rangle_{v_0}}\,.
}
Thus, we may rewrite the bulk-to-boundary propagator as
\eqn{Kbdy}{
\hat{K}_{\Delta}(z,x) = \frac{\psi_\Delta(\tilde{q}_{{\rm join}(x,u,v)})}{\psi_\Delta(\tilde{q}_{v_0})} \left(\frac{|u-v|_\mathfrak{q}}{|x-u|_\mathfrak{q} |x-v|_\mathfrak{q}} \right)^\Delta,
}
for the configuration depicted in Fig.~\ref{fig:Kxuv}. Up to overall homogeneity degree-dependent factors, this matches the expression on the regular tree~\cite{Gubser:2016guj}.\footnote{On the regular tree, an additional equivalent form of the bulk-to-boundary propagator exists, more closely resembling the continuum AdS expression in Poincar\'{e} coordinates~\cite{Gubser:2016guj}. 
From the $p$-adic perspective, it involves a parametrisation of bulk points in terms of a boundary coordinate and a bulk ``depth,'' which is defined with respect to a chosen base point $v_0$ belonging to the boundary.
A similar construction is possible on the biregular tree and can be generalised to accommodate base vertices in the bulk. This form of the bulk-to-boundary propagator is not pertinent to our discussion here, so we omit it for brevity.} 
Finally, it is obvious from the above that $\hat{K}_\Delta$ is obtained from a limiting procedure of $\hat{G}_\Delta$, as follows,
\eqn{GtoK}{
\hat{K}_\Delta(a,x_3) = \lim_{x_4 \to x_3} \frac{\psi_{\Delta}({q}_{{\rm join}(x_3,x_4,x_1)})}{\psi_\Delta(\tilde{q}_{v_0})} |x_{34}|_\mathfrak{q}^{-\Delta} \hat{G}_\Delta(a,b)\,.
}

\subsection{Two-point function: Approaching the boundary}
\label{sec:TWOPT}

To compute the two-point function of the dual (boundary) theory with the correct normalisation, the unambiguous approach is to evaluate the on-shell quadratic action to obtain a regularised boundary term contribution, from which the two-point function can be extracted~\cite{Gubser:1998bc}. 
The normalisation constant is most reliably extracted in Fourier space.
An analogous procedure was employed for the ($p$+1)-regular tree action (for $p$ prime) to compute the two-point function in the dual $p$-adic CFT~\cite{Gubser:2016guj}.

The boundary of the biregular tree is more intricate and technically challenging. We defer the discussion of Fourier analysis and the precise computation of the two-point function by evaluating the on-shell action for the future. In this section, we present an alternative perspective, along the lines of Ref.~\cite{Heydeman:2016ldy}.

Naively, one can obtain the two-point function by taking a limit of the bulk-to-boundary propagator as the bulk vertex approaches the boundary. 
In Fig.~\ref{fig:Kxuv}, this corresponds to boundary points $u$ and $v$ approaching each other and limiting to boundary point $y$, which also sends bulk vertex $z$ to the boundary.
Taking the limit in~\eqref{Kbdy}, we get
\eqn{}{
\langle {\cal O}_{\Delta}(x) {\cal O}_{\Delta}(y) \rangle &\sim \lim_{u,v \to y} 
\frac{\psi_\Delta(\tilde{q}_{v_0}) |u-v|_\mathfrak{q}^{-\Delta}} {\psi_{\Delta}(\tilde{q}_{{\rm join}(x,u,v)})}
\hat{K}_{\Delta}(z,x) \cr 
&= \frac{1}{|x-y|_\mathfrak{q}^{2\Delta}},
}
up to overall factors which cannot be fixed with this limiting procedure.
In terms of the bulk-to-bulk propagator~\eqref{Gbdy}, this corresponds to
\eqn{}{
& \langle {\cal O}_{\Delta}(x_1) {\cal O}_{\Delta}(x_3) \rangle \cr 
&\sim \lim_{\substack{x_2 \to x_1\\ x_4 \to x_3}} \frac{\psi_\Delta({q}_{{\rm join}(x_3,x_4,x_1)}) } {\psi_{\Delta}(\tilde{q}_{{\rm join}(x_1,x_2,x_3)})} 
|x_{12}|_\mathfrak{q}^{-\Delta} |x_{34}|_\mathfrak{q}^{-\Delta}
\hat{G}_{\Delta}(a,b) \cr 
&= \frac{1}{|x_{13}|_\mathfrak{q}^{2\Delta}}\,.
}
The overall factors of $\psi_\Delta$s are necessary so that the boundary limit is well-defined and does not depend on whether the bulk points $a$ and $b$ are sent to the boundary while keeping $d(a,b)$ even or odd.
The scaling of the two-point function is consistent with the two-point function in a dual CFT on the boundary. 

In the rest of this section, the letter $v$ will be reserved to denote a bulk vertex.
We now write down the general bulk solution, $\phi(v)$, to the equation of motion on the biregular tree.  
It is given by a linear superposition of primitive solutions subject to Dirichlet boundary conditions, $\phi_0(x)$ (with appropriate scaling as $v \rightarrow x$ on the boundary),
\eqn{harmonic}{
\phi(v) = N \int_{\partial T_{q_+,q_-}} \!\!\!\!\!\! d\mu_0(x) \phi_0(x) 
\frac{\psi_{\Delta}(\tilde{q}_v)}{\psi_{\Delta}(\tilde{q}_{v_0})} \mathfrak{q}^{\Delta \langle v,x \rangle_{v_0}} \,,
}
\replaced{where $N$ is a normalization constant as yet
unspecified.}{up to some normalisation to be fixed.} Here $\mu_0$ is the Patterson-Sullivan measure on the boundary $\partial T_{q_+,q_-}$, defined as follows.

 Let $B_u \subseteq T_{q_+,q_-}$ be the set of all bulk vertices that can be reached by geodesic rays starting at base vertex $v_0$ and passing through $u \in T_{q_+,q_-}$. 
Let $\partial B_u \subseteq \partial T_{q_+,q_-}$ be the set of all {\it boundary} points that can be reached by geodesic rays starting at $v_0$ and passing through $u$.
Define the measure $\mu_0$ to be,
\eqn{}{
\mu_{0}(\partial B_u) := \begin{cases}
    q_{v_0}^{-1} \mathfrak{q}^{1-d(v_0,u)} & d(v_0,u) {\rm \ odd} \cr 
    \mathfrak{q}^{-d(v_0,u)} & d(v_0,u) {\rm \ even}
\end{cases}\,.
}
The measure is normalized such that $\mu_{0}(\partial T_{q_+,q_-}) = (q_{v_0}+1)/q_{v_0}$.\footnote{The measure we use differs in normalisation compared to the {\it equidistributed boundary measure} used in Ref.~\cite{tarabusi2022}, but in the regular tree limit has the right reduction to the Patterson-Sullivan measure on regular trees~\cite{Heydeman:2016ldy}.}

To work out the behaviour of the harmonic function~\eqref{harmonic} near the boundary, we choose $\phi_0(x)$ to be the characteristic function of an open ball on the boundary, $\partial B_w$, for some $w \in T_{q_+,q_-}$. Then,
\eqn{}{
\phi_w(v) := N \frac{\psi_{\Delta}(\tilde{q}_v)}{\psi_{\Delta}(\tilde{q}_{v_0})} \int_{\partial B_w} \!\!\! d\mu_0(x) 
\mathfrak{q}^{\Delta \langle v,x \rangle_{v_0}} \,.
}
When $v \notin B_w$, then $\langle v,x\rangle_{v_0} = \langle v,w \rangle_{v_0}$ and so is independent of $x$. Thus, the integral evaluates to
\eqn{}{
\phi_w(v) = N \frac{\psi_{\Delta}(\tilde{q}_v)}{\psi_{\Delta}(\tilde{q}_{v_0})} \mathfrak{q}^{\Delta \langle v,w \rangle_{v_0}} \mu_0(\partial B_w) 
 \,.
}
For $v \in B_w$, we must consider two possibilities: (i) $x \in \partial B_v$, or (ii) $x \notin \partial B_v$. In the first case, $\langle v,x \rangle_{v_0} = d(v_0,v)$ and the integral becomes
\eqn{poss1}{
\phi_w(v)|_{(i)} = N \frac{\psi_{\Delta}(\tilde{q}_v)}{\psi_{\Delta}(\tilde{q}_{v_0})} \mathfrak{q}^{\Delta d(v_0,v)} \mu_0(\partial B_v) \,.
}
For the second possibility, the vertex $j:={\rm join}(v,x,v_0)$ belongs to the geodesic segment $(v:w]$ such that if $h:=d(v,j)$, then as $x$ varies, $h$ varies over the range $\{1,2,\ldots, d(v,w)\}$ (see Fig.~\ref{fig:AppBdy}). 
\begin{figure}
    \centering
\begin{tikzpicture}[scale=1]
    \draw (0,0) circle (2.5);
    
    \coordinate (x) at (60:2.5);
    \coordinate (w) at (-90:0.5);
    \coordinate (v0) at (-90:1.5);
    \coordinate (v) at (90:1.8);
    \coordinate (j) at (90:0.4);
    \coordinate (jm) at (90:0.9);
    \coordinate (branch1) at (0:0.5);
    \coordinate (branch2) at (180:0.5);
    
    \draw[thick] (v0) -- (w);
    
    \draw[thick] (w) -- (j) -- (x);
    
    \draw[thick] (w) -- (branch1);
    \draw[thick] (w) -- (branch2);
    
    \draw[thick] (branch2) -- (-150:1.2);
    \draw[thick] (branch2) -- (-210:1.2);
    \draw[thick] (-150:1.2) -- (-135:2.5);
    \draw[thick] (-150:1.2) -- (-165:2.5);
    \draw[thick] (-210:1.2) -- (-195:2.5);
    \draw[thick] (-210:1.2) -- (-225:2.5);
    
    \draw (-135:2.3) -- (-130:2.5);
    \draw (-135:2.3) -- (-140:2.5);
    \draw (-165:2.3) -- (-160:2.5);
    \draw (-165:2.3) -- (-170:2.5);
    \draw (-195:2.3) -- (-190:2.5);
    \draw (-195:2.3) -- (-200:2.5);
    \draw (-225:2.3) -- (-220:2.5);
    \draw (-225:2.3) -- (-230:2.5);
    
    \draw[thick] (branch1) -- (30:1.2);
    \draw[thick] (branch1) -- (-30:1.2);
    \draw[thick] (30:1.2) -- (15:2.5);
    \draw[thick] (30:1.2) -- (45:2.5);
    \draw[thick] (-30:1.2) -- (-15:2.5);
    \draw[thick] (-30:1.2) -- (-45:2.5);
    
    \draw (15:2.3) -- (10:2.5);
    \draw (15:2.3) -- (20:2.5);
    \draw (45:2.3) -- (40:2.5);
    \draw (45:2.3) -- (50:2.5);
    \draw (-15:2.3) -- (-10:2.5);
    \draw (-15:2.3) -- (-20:2.5);
    \draw (-45:2.3) -- (-40:2.5);
    \draw (-45:2.3) -- (-50:2.5);

    \draw (60:2.3) -- (65:2.5);
    \draw (60:2.3) -- (55:2.5);
    
    \draw[thick] (j) -- (90:1.8);
    \draw[thick] (90:1.8) -- (85:2.5);
    \draw[thick] (90:1.8) -- (95:2.5);
    \draw (85:2.3) -- (82:2.5);
    \draw (85:2.3) -- (88:2.5);
    \draw (95:2.3) -- (92:2.5);
    \draw (95:2.3) -- (98:2.5);

    \draw[red, thick, decoration={brace, amplitude=7pt}, decorate] 
        (v) -- (j) node[midway, red, right=4pt] {$h$};
    
    \fill[red] (x) circle (2pt);
    \fill (v0) circle (2pt);
    \fill (w) circle (2pt);
    \fill (j) circle (2pt);
    \fill (jm) circle (2pt);
    \fill (v) circle (2pt);
    
    \node[red, above right] at (x) {\large $x$};
    \node[below] at (v0) {\large $v_0$};
    \node[below left] at (w) {\large $w$};
    \node[left] at (v) {\large $v$};
    \node[left] at (j) {\large $j$};
    \node[left] at (jm) {\large $\tilde{j}_-$};

\end{tikzpicture}
    \caption{Configuration on the biregular tree for $v \in B_w$ with $x \notin \partial B_v$.}
\label{fig:AppBdy}

\end{figure}
In this case, $\langle v,x \rangle_{v_0} = d(v_0,v)-2h$.
For each value of $h$, the integrand is constant over the \added{set $\partial B_h := \partial B_j \setminus \partial B_{\widetilde{j}_-}$ where $\tilde{j}_-$ represents the nearest neighbour of $j$ closest to $v$ (see Figs.~\ref{fig:AppBdy}-\ref{fig:Telescope})}. \added{This set has measure,} 
\eqn{}{
\mu_0(\partial B_h) = \begin{cases}
    q_{v_0}^{-1} \frac{(q_j-1)}{q_j}\mathfrak{q}^{1-(d(v_0,v)-h)}  & d(v,v_0)-h {\rm \ odd} \cr 
    \frac{(q_j-1)}{q_j} \mathfrak{q}^{-(d(v,v_0)-h)} & d(v,v_0)-h {\rm \ even}
\end{cases}
}
\deleted{where we abbreviated ${\rm join}(v,x,v_0) =:j$.} 
Note that, 
\eqn{}{ 
\mu_0(\partial B_w) = \mu_0(\partial B_v) + \sum_{h=1}^{d(v,w)} \mu_0(\partial B_h)\,,
}
\added{which is easily proven upon recognising that the right-hand-side leads to a telescoping series following from the definition of $\partial B_h$.}
\begin{figure}[t]
\centering
\begin{tikzpicture}[scale=1]

\draw[thick] (0,0) circle (2.5);
\draw[thick] (0,0) circle (1.6);
\draw[thick] (0,0) circle (1.2);
\draw[thick] (0,0) circle (.5);
  
  \node at (0,0) { $\partial B_v$};           
  \node at (0,-0.8) { $\partial B_{\widetilde{j}_-}$};  
  \node at (0,-1.4) { $\partial B_j$};       
  \node at (0,-2.0) { $\partial B_w$};       
  
  \fill[red!90] (0.7,-1.25) circle (2pt);
  \node[red!90] at (0.95,-1.05) { $x$};
  
\end{tikzpicture}
\caption{A schematic representation of set inclusion for the configuration shown in Fig.~\ref{fig:AppBdy}.}
\label{fig:Telescope}

\end{figure}

Then the contribution to the harmonic function from the second possibility becomes
\eqn{poss2}{
\phi_w(v)|_{(ii)} = N \frac{\psi_{\Delta}(\tilde{q}_v)}{\psi_{\Delta}(\tilde{q}_{v_0})}  
\mathfrak{q}^{\Delta d(v_0,v)} \sum_{h=1}^{d(v,w)} \mathfrak{q}^{-2h\Delta} \mu_0(\partial B_h) \,.
}
\begin{widetext}
\noindent Adding up~\eqref{poss1} and~\eqref{poss2}, with some work we obtain
\eqn{harmonicBw}{
\phi_w(v) &=  \left(N \frac{\zeta^-_\mathfrak{q}(2\Delta-1)}{\zeta^-_\mathfrak{q}(2\Delta)}  \right)
\sqrt{\frac{q_v}{q_{v_0}}}
\frac{\tilde{q}_v + \mathfrak{q}^{2\Delta}}{\mathfrak{q} + \mathfrak{q}^{2\Delta}}
\frac{\psi_{\Delta}(\tilde{q}_v)}{\psi_{\Delta}(\tilde{q}_{v_0})}  
\mathfrak{q}^{(\Delta-1) d(v_0,v)}  \cr 
&-\left(N \sqrt{\frac{q_w}{q_{v_0}}} 
\frac{\mathfrak{q}^{(2\Delta-1) d(v_0,w)}}{\mathfrak{q}^{2\Delta}-\mathfrak{q}} 
\frac{\tilde{q}_w (q_w-1) + \mathfrak{q}^{2\Delta} (\tilde{q}_w-1) }{\mathfrak{q}+\mathfrak{q}^{2\Delta}} \right)
\frac{\psi_{\Delta}(\tilde{q}_v)} {\psi_{\Delta}(\tilde{q}_{v_0})}  
\mathfrak{q}^{-\Delta d(v_0,v)} \,,
}
for $v \in B_w$.
\end{widetext}
\added{In writing~\eqref{harmonicBw}, we utilized the local zeta function}\footnote{\added{The non-standard superscript in $\zeta_\mathfrak{q}^-$ is introduced for future convenience, as in the next section we shall also define a related local zeta function, $\zeta_\mathfrak{q}^+$.} }
\eqn{LocalZetaMinus}{
\zeta_{\mathfrak{q}}^{-}(s) := \frac{1}{1 - \mathfrak{q}^{-s}}\,.
}
It can be checked that both terms in~\eqref{harmonicBw} are individually harmonic and satisfy the bulk equation of motion. 
Moreover,~\eqref{harmonicBw} reproduces the regular tree result~\cite{Heydeman:2016ldy} in the appropriate limit.

In the limit when the bulk point $v$ approaches the boundary point $x \in \partial B_v \subseteq \partial B_w$, the leading behaviour comes from the first term in~\eqref{harmonicBw} as $d(v_0,v) \rightarrow \infty$, since we are assuming $\Delta > 1/2$.
Thus, setting 
\eqn{NChoice}{ 
N = \frac{\zeta^-_\mathfrak{q}(2\Delta)}{\zeta^-_\mathfrak{q}(2\Delta-1)} \frac{\mathfrak{q}+\mathfrak{q}^{2\Delta}}{\tilde{q}_{v_0}+\mathfrak{q}^{2\Delta}}
}
and letting $v \rightarrow x\in \partial B_v \subseteq \partial B_w$ in~\eqref{harmonic}, we get
\eqn{asympt}{
\phi(v) \sim \sqrt{\frac{q_v}{q_{v_0}}}
\frac{\tilde{q}_v + \mathfrak{q}^{2\Delta}}{\tilde{q}_{v_0} + \mathfrak{q}^{2\Delta}}
\frac{\psi_{\Delta}(\tilde{q}_v)}{\psi_{\Delta}(\tilde{q}_{v_0})}  
\mathfrak{q}^{(\Delta-1) d(v_0,v)} \phi_0(x)\,.
}
This is the moral equivalent of the boundary limit of the bulk scalar field in continuum AdS${}_2$/CFT${}_1$,
\eqn{}{
\lim_{z\to 0} \phi(z,\vec{x}) \sim z^{1-\Delta} \phi_0(\vec{x})
}
with $\mathfrak{q}^{-d(v_0,v)}$ identified with a (normalized) bulk ``depth'' coordinate $z$, approaching smaller and smaller values as the bulk point $v$ moves farther and farther from the base point $v_0$. 
The $q_v$-dependent prefactors in~\eqref{asympt} are needed to ensure that $\phi(v)$ satisfies the bulk equation of motion.
If we arrange to send $v \to x$ in steps of two units such that $d(v_0,v)$ is always even, the degree-dependent factors in~\eqref{asympt} cancel, leaving 
\eqn{asymptEven}{
\phi(v) \sim 
\left(\mathfrak{q}^{-d(v_0,v)}\right)^{1-\Delta}\phi_0(x) \qquad d(v_0,v) = 2k \,, k \to \infty\,.
}
\added{We note that the choice of $N$ is non-unique. For any choice, the prefactor to $\mathfrak{q}^{(\Delta-1)d(v_0,v)} \phi_0(x)$ in~\eqref{asympt} alternates as the parity of $d(v_0,v)$ alternates, a peculiar feature of the semihomogeneous space, except in the massless limit, where we obtain the (unique) limiting value $\phi(v) \sim \phi_0(x)$ as $v \to x$, for $N$ given by~\eqref{NChoice}.}

Let $y \in \partial B_v$ be another boundary point such that the Gromov product of $x$ and $y$ is $(x,y)_{v_0} = d(v_0,v)$.\footnote{This is the same as requiring ${\rm join}(x,y,v_0) = v$.}
Then, using the visual metric~\eqref{VisualMetric} to measure  boundary distances, we have the equality, $|x-y|_\mathfrak{q} = \mathfrak{q}^{-d(v_0,v)}$.
The quantity $|x-y|_\mathfrak{q}$ quantifies bulk depth, getting smaller as we get closer to the boundary.
Using this,~\eqref{harmonic} can be rewritten as the following boundary integral, a precursor to the extrapolate dictionary~\cite{Balasubramanian:1998de,Banks:1998dd,Harlow:2011ke}
\eqn{harmonicAgain}{
\phi(v) &= \frac{\zeta^-_\mathfrak{q}(2\Delta)}{\zeta^-_\mathfrak{q}(2\Delta-1)} \frac{\mathfrak{q}+\mathfrak{q}^{2\Delta}}{\tilde{q}_{v_0}+\mathfrak{q}^{2\Delta}}
\frac{\psi_{\Delta}(\tilde{q}_v)}{\psi_{\Delta}(\tilde{q}_{v_0})}\,
\mathfrak{q}^{-\Delta d(v_0,v)}\cr 
&\times \int_{\partial T_{q_+,q_-}} \!\!\!\!\!\! d\mu_0(x) \frac{\phi_0(x)}{ |x-y|_\mathfrak{q}^{2\Delta}} \,,
}
where we used $\langle v,x\rangle_{v_0} = d(v_0,v) = -d(v_0,v) - 2 \log_{\mathfrak{q}}|x-y|_\mathfrak{q}$.

One could, in principle, proceed to define a nonlocal boundary derivative operator on the boundary of the biregular tree, analogous to the Vladimirov derivative (see, {\it e.g.}, Ref.~\cite{Zabrodin1989} and more recent work~\cite{bradley2025}).
Indeed, it would be interesting to start from the local bulk action on the biregular tree and derive a nonlocal boundary action for the dual theory, along the lines of Ref.~\cite{Zabrodin1989}.
We leave this and many more questions involving nonlocal ultrametric analysis on the fractal boundary~\cite{bradley2025} for future work.

\section{Three-point correlator}
\label{sec:THREEPT}

In this section, we calculate the three-point contact diagram, represented by the bulk integral
\eqn{A3Def}{
A_3(x_1,x_2,x_3) := \!\!\!\!\!\sum_{z \in T_{q_+,q_-}} \!\!\!\!\! \hat{K}_{\Delta_1}(z,x_1) \hat{K}_{\Delta_2}(z,x_2) \hat{K}_{\Delta_3}(z,x_3) \,,
}
where $x_1, x_2, x_3 \in \partial T_{q_+,q_-}$ are boundary points. 
Such a contribution can arise in a theory of three massive scalars of generic scaling dimensions $\Delta_1, \Delta_2$, and $\Delta_3$ with a cubic interaction vertex of the form $\phi^{(\Delta_1)}_a\phi^{(\Delta_2)}_a\phi^{(\Delta_3)}_a$. 
For instance, for a single self-interacting scalar field of dual scaling dimension $\Delta$, the action  takes the form
\eqn{Scubic}{
S[\phi] &=  \sum_{\langle ab \rangle} \frac{1}{ 2} (\phi_a - \phi_b)^2 + \sum_{a \in T_{q_+,q_-}} \!\!\!\left( \frac{1}{2} m_{\Delta}^2 \phi_a^2 + \frac{g_3}{3!} \phi_a^3 \right).
}

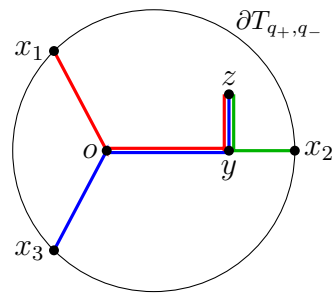
\begin{figure}[t]
\centering
\begin{tikzpicture}[scale=1.25]
  \draw (0,0) circle (1.5);
  
  \coordinate (x1) at (135:1.5);
  \coordinate (x2) at (0:1.5);
  \coordinate (x3) at (225:1.5);
  
  \coordinate (c) at (-0.5,0);
  
  \coordinate (b1) at (0.75,0);
  \coordinate (a1) at (0.75,0.6);
  \coordinate (b2) at (0.8,0);
  \coordinate (a2) at (0.8,0.6);
  \coordinate (b3) at (0.85,0);
  \coordinate (a3) at (0.85,0.6);
  
  \draw[red, very thick] (x1) -- (c) -- (c |- 0,0.025) -- (b1 |- 0,0.025) -- (b1) -- (a1);
  
  \draw[blue, very thick] (x3) -- (c) -- (c |- 0,-0.02) -- (b2 |- 0,-0.02) -- (b2) -- (a2);
  
  \draw[green!70!black, very thick] (x2) -- (b3) -- (a3);
  
  \fill (x1) circle (1.5pt) node[left] {\large $x_1$};
  \fill (x2) circle (1.5pt) node[right] {\large $x_2$};
  \fill (x3) circle (1.5pt) node[left] {\large $x_3$};
  
  \fill (c) circle (1.5pt) node[left] {\large $o$};
  \fill (b2) circle (1.5pt) node[below] {\large $y$};
  \fill (a2) circle (1.5pt) node[above] {\large $z$};
  \node at (45:1.9) { $\partial T_{q_+,q_-}$};
\end{tikzpicture}
  \caption{Subway diagram~\cite{Gubser:2016guj} for the product $\prod_{i=1}^3 \hat{K}_{\Delta_i}(z,x_i)$, showing geodesics from boundary points $x_1, x_2, x_3$ to bulk point $z$. All geodesics are paths on the biregular tree. The geodesics joining all boundary points intersect at the vertex $o={\rm join}(x_1,x_2,x_3)$ on the biregular tree.}\label{fig:ThreePt}
 \end{figure}

First, we will make some general remarks on the structure of the three-point function. On a tree geometry, geodesics joining three non-coincident boundary points meet at a unique bulk point of intersection, denoted $o \in T_{q_+,q_-}$, {\it i.e.} $o={\rm join}(x_1,x_2,x_3)$ (see Fig.~\eqref{fig:ThreePt}).
The three-point correlator cannot depend on any fixed bulk vertex $o$ in homogeneous spaces, such as a regular tree. 
For instance, on a $(p+1)$-regular tree, in the context of $p$-adic AdS/CFT, one can employ $p$-adic linear fractional transformations to map the bulk point $o$ to any other bulk point. 
However, on a biregular tree, a semihomogeneous space with only the subgroup of even translations permitted as part of bulk isometries, the vertex $o$ can only be mapped to other vertices of the same degree. 
Consequently, the degree of vertex $o$, denoted $q_o$, emerges as an invariant.\footnote{The non-trivial dependence of the propagators~\eqref{Gbdy} and~\eqref{Kbdy} on $\psi_\Delta(\cdot)$ is precisely for the same reason.}
As we will demonstrate, this leads to a nontrivial dependence of the three-point correlator on $q_o$ and the emergence of a nontrivial ``tensor structure'' constructed out of $q_o$. 

To explicitly evaluate~\eqref{A3Def}, we partition the total bulk sum into four partial sums, corresponding to whether the bulk vertex $z$ branches out from a bush rooted at any vertex $y \in (o:x_i)$ for $i=1,2,3$, and if not, when it branches out of $y=o$. 
In each case, we rewrite the summand using the forwards/backwards splitting identity, proven in App.~\ref{APP:SPLIT},\footnote{This identity is essentially the same as the one on the regular tree, except for the inclusion of zero-displacement propagator factors such as $\hat{G}_\Delta(w,w)$ that are no longer constant on the biregular tree, but rather depend on the vertex’s degree.}
\eqn{Ksplit}{
\hat{K}_\Delta(z,x) = \begin{cases} \displaystyle{\frac{\hat{K}_\Delta(w,x)\hat{G}_\Delta(w,z)}{\hat{G}_{\Delta}(w,w)}}  & w \in (x,z] \cr 
\displaystyle{\frac{\hat{K}_\Delta(w,x) \hat{G}_{\Delta}(z,z)}{\hat{G}_\Delta(w,z)}}  & z \in (x,w] 
\end{cases}\,.
}
For example, when $y \in (o:x_2)$ (see the configuration depicted in Fig.~\ref{fig:ThreePt}), we have 
\eqn{}{
\hat{K}_{\Delta_1}(z,x_1) &= \frac{\hat{K}_{\Delta_1}(o,x_1)\hat{G}_{\Delta_1}(o,z)}{\hat{G}_{\Delta_1}(o,o)} \cr 
\hat{K}_{\Delta_2}(z,x_2) &= \frac{\hat{K}_{\Delta_2}(y,x_2)\hat{G}_{\Delta_2}(y,z)}{\hat{G}_{\Delta_2}(y,y)} \cr 
\hat{K}_{\Delta_3}(z,x_3) &= \frac{\hat{K}_{\Delta_3}(o,x_3)\hat{G}_{\Delta_3}(o,z)}{\hat{G}_{\Delta_3}(o,o)}
}
and we also have
\eqn{}{
\hat{K}_{\Delta_2}(y,x_2) &= \frac{\hat{K}_{\Delta_2}(o,x_2)\hat{G}_{\Delta_2}(y,y)}{\hat{G}_{\Delta_2}(o,y)} \,.
}
Substituting these in~\eqref{A3Def}, the summand becomes
\eqn{}{
\frac{\left( \prod_{i=1}^3 \hat{K}_{\Delta_i}(o,x_i)\right)}{\hat{G}_{\Delta_1}(o,o) \hat{G}_{\Delta_3}(o,o)} \frac{\hat{G}_{\Delta_1}(o,z) \hat{G}_{\Delta_2}(y,z) \hat{G}_{\Delta_3}(o,z)}{\hat{G}_{\Delta_2}(o,y)} \,,
}
for $y \in (o:x_2)$.
\replaced{It remains to carry out the sum over all possible vertices $y \in (o:x_2)$ and the corresponding $z$. To do this, in an intermediate step, we partition the sum based on vertex type to obtain multiple simpler geometric series.}{It is now straightforward to sum over all possible vertices $y \in (o:x_2)$ and the corresponding $z$.}
Evaluating the resulting geometric series, we obtain, up to an overall factor of $\left( \prod_{i=1}^3 \hat{K}_{\Delta_i}(o,x_i)\right) $,\footnote{For the geometric series to converge, we assume $\Delta_{1}+\Delta_3-\Delta_2>0$. Likewise, in evaluating contributions from the other legs, we will need to assume $\Delta_{2}+\Delta_3-\Delta_1>0$ and $\Delta_{1}+\Delta_2-\Delta_3>0$.}  
\eqn{}{
&- \left[ \frac{\mathfrak{q}^{-2\Delta_2}}{1-\mathfrak{q}^{2\Delta_{13,2}}}\Psi_{123}(\tilde{q}_o)^2 \!\left( 1+ \frac{(\tilde{q}_o-1)(q_o+\Psi_{123}(q_o)^2)}{-\mathfrak{q}^2+\mathfrak{q}^{2\Delta_{123,}}}\right) \right. \cr 
& \left. + \frac{1}{1-\mathfrak{q}^{2\Delta_{13,2}}} \left( 1 + \frac{(q_o-1)(\tilde{q}_o+\Psi_{123}(\tilde{q}_o)^2)}{-\mathfrak{q}^2+\mathfrak{q}^{2\Delta_{123,}}}\right)\right],
}
where we defined
\eqn{}{
\Psi_{123}(q_o) &:=  \frac{1}{\psi_{\Delta_1}(q_o)\psi_{\Delta_2}(q_o)\psi_{\Delta_3}(q_o) } \,,
}
so that the following identity holds,
\eqn{}{
\Psi_{123}(q_o) \Psi_{123}(\tilde{q}_o) = \mathfrak{q}^{\Delta_{123,}}\,.
}
Here and below, we use the notation
\eqn{Deltaijk}{
\Delta_{ij,k} := \Delta_i + \Delta_j - \Delta_k \qquad 
\Delta_{ijk,} := \Delta_i + \Delta_j + \Delta_k \,.
}
The contributions from the legs $(o:x_1)$ and $(o:x_3)$ can be obtained from the above using cyclic symmetry. The remaining contribution from $y=o$ is also straightforward to evaluate, yielding
\eqn{}{
&\left( \prod_{i=1}^3 \hat{K}_{\Delta_i}(o,x_i)\right) \left[ 1 + \frac{(q_o-2)(\tilde{q}_o+\Psi_{123}(\tilde{q}_o)^2)}{-\mathfrak{q}^2+\mathfrak{q}^{2\Delta_{123,}}} \right].
}

All contributions are proportional to the same product of bulk-to-boundary propagators from the boundary points to the join of the boundary points. This factor can be rewritten in terms of the visual metric~\eqref{VisualMetric} on the boundary as follows,
\eqn{KKKbdy}{
\left( \prod_{i=1}^3 \hat{K}_{\Delta_i}(o,x_i)\right) = \frac{  \Psi_{123}(\tilde{q}_{v_0})/\Psi_{123}(\tilde{q}_{o}) }{|x_{12}|_\mathfrak{q}^{\Delta_{12,3}} |x_{23}|_\mathfrak{q}^{\Delta_{23,1}} |x_{31}|_\mathfrak{q}^{\Delta_{31,2}}}\,.
}
To prove~\eqref{KKKbdy}, we assume, without loss of generality, the configuration of boundary points depicted in Fig.~\ref{fig:ultrametric}.
Then, using~\eqref{Kbdy}, it is straightforward to work out the propagators,
\eqn{}{
\hat{K}_{\Delta_1}(o,x_1) &= \frac{\psi_{\Delta_1}(\tilde{q}_o)}{\psi_{\Delta_1}(\tilde{q}_{v_0})} 
\left(\frac{|x_{23}|_\mathfrak{q}}{|x_{12}|_\mathfrak{q} |x_{13}|_\mathfrak{q}}\right)^{\Delta_1}
 \cr 
\hat{K}_{\Delta_2}(o,x_2) 
&= 
\frac{\psi_{\Delta_2}(\tilde{q}_o)}{\psi_{\Delta_2}(\tilde{q}_{v_0})} 
\left(\frac{|x_{13}|_\mathfrak{q}}{|x_{12}|_\mathfrak{q} |x_{23}|_\mathfrak{q}}\right)^{\Delta_2}
 \cr 
\hat{K}_{\Delta_3}(o,x_3) 
&= \frac{\psi_{\Delta_3}(\tilde{q}_o)}{\psi_{\Delta_3}(\tilde{q}_{v_0})}  
\left(\frac{|x_{12}|_\mathfrak{q}}{|x_{13}|_\mathfrak{q} |x_{23}|_\mathfrak{q}}\right)^{\Delta_3} \,,
}
from which~\eqref{KKKbdy} follows immediately.

Adding all contributions together and reorganising the terms, we obtain,
\eqn{A3val}{
A_3(x_1,x_2,x_3) &= f_{123}^-\, \frac{V_{123}^{+}(q_o)}{|x_{12}|_\mathfrak{q}^{\Delta_{12,3}} |x_{23}|_\mathfrak{q}^{\Delta_{23,1}} |x_{31}|_\mathfrak{q}^{\Delta_{31,2}}} \cr 
&+ f_{123}^+\, \frac{V_{123}^{-}(q_o)}{|x_{12}|_\mathfrak{q}^{\Delta_{12,3}} |x_{23}|_\mathfrak{q}^{\Delta_{23,1}} |x_{31}|_\mathfrak{q}^{\Delta_{31,2}}}\,,
}
where the precursors to ``tensor structures'' for the three-point function on the biregular tree take the form
\eqn{V123Def}{
&V_{123}^{\pm}(q_o) := \frac{\Psi_{123}(\tilde{q}_{v_0})}{2(\mathfrak{q} \pm \mathfrak{q}^{\Delta_{123,}})}  \!\left(     \frac{\tilde{q}_o \pm \mathfrak{q}^{\Delta_{123,}}}{\Psi_{123}(\tilde{q}_o)}  \pm  \frac{q_o \pm \mathfrak{q}^{\Delta_{123,}}}{\Psi_{123}(q_o)}\!\right)\!.
}   
The constant coefficients appearing in \eqref{A3val} are given by
\eqn{f123Def}{
 f^\pm_{123}  &= \frac{ \zeta^\pm_{\mathfrak{q}}(\Delta_{{123,}}-1) \zeta^\pm_{\mathfrak{q}}(\Delta_{12,3}) \zeta^\pm_{\mathfrak{q}}(\Delta_{23,1}) \zeta^\pm_{\mathfrak{q}}(\Delta_{31,2}) }{ \zeta^-_{\mathfrak{q}}(2\Delta_1) \zeta^-_{\mathfrak{q}}(2\Delta_2) \zeta^-_{\mathfrak{q}}(2\Delta_3) }\,,
}
which are expressed in terms of the local zeta functions
\eqn{LocalZeta}{
\zeta_{\mathfrak{q}}^{\pm}(s) := \frac{1}{1 \pm \mathfrak{q}^{-s}}\,.
}
\added{We will comment on the appearance of the two types of local zeta functions above in Sec.~\ref{sec:DISCUSS}.}

Observing that $V_{123}^\pm(q_o) = \pm V_{123}^\pm(\tilde{q}_o)$, we may simplify the $q_o$ dependence in the three-point function by writing
\eqn{V123Again}{
V_{123}^\pm(q_o) = (\pm 1)^{\delta_{q_o,q_+}} \Psi_{123}(\tilde{q}_{v_0}) W_{123}^\pm \,,
}
 where $\delta_{q_a,q_b}$ is a Kronecker delta for homogeneity degrees, and we have defined bulk vertex-independent constants
\eqn{W123Def}{
W_{123}^\pm := \frac{1}{2(\mathfrak{q} \pm \mathfrak{q}^{\Delta_{123,}})} 
\left( \frac{{q}_+ \pm \mathfrak{q}^{\Delta_{123,}}}{\Psi_{123}({q}_+)}  \pm  \frac{q_- \pm \mathfrak{q}^{\Delta_{123,}}}{\Psi_{123}(q_-)}\right).
}
Thus, we arrive at the final form,\footnote{The homogeneity degree $q_o$ may also be written as $q_{{\rm join}(x_1,x_2,x_3)}$ directly in terms of the boundary data.} 
\eqn{ThreePt}{
A_3(x_i) &= f_{123}^- \frac{ W^+_{123} \Psi_{123}(\tilde{q}_{v_0})}{|x_{12}|_\mathfrak{q}^{\Delta_{12,3}} |x_{23}|_\mathfrak{q}^{\Delta_{23,1}} |x_{31}|_\mathfrak{q}^{\Delta_{31,2}}} \cr 
&+   f_{123}^+  (-1)^{\delta_{q_o,q_+}}\frac{ W^-_{123} \Psi_{123}(\tilde{q}_{v_0})}{ |x_{12}|_\mathfrak{q}^{\Delta_{12,3}} |x_{23}|_\mathfrak{q}^{\Delta_{23,1}} |x_{31}|_\mathfrak{q}^{\Delta_{31,2}}}\,.
}
Here, the two terms can be interpreted as analogues of invariant three-point ``tensor structures'' on the biregular tree.

The dependence on the degree of the base vertex $q_{v_0}$, appearing as an overall factor in~\eqref{ThreePt}, arises from the choice of normalisation of the bulk-to-boundary propagator~\eqref{Khat}.
Physically meaningful, invariant OPE coefficients are extracted by dividing the three-point function by an appropriate power of products of two-point functions.
As previously mentioned, in the absence of a correctly normalised two-point function, we leave such considerations for the future.

Upon setting $q_a=\tilde{q}_a = \mathfrak{q}$ for all bulk points $a$, the three-point result manifestly reduces to the three-point function on a $(\mathfrak{q}+1)$-regular tree since clearly, $W_{123}^-$  vanishes identically and $W^+_{123}=1/\Psi_{123}(\mathfrak{q})$, so that the structure reduces to the well-known form for a scalar three-point function in a ($\mathfrak{q}$-adic) CFT~\cite{Melzer1989},
\eqn{A3regular}{
A_3^{\rm (regular)}(x_1,x_2,x_3) = 
   \frac{{f}_{123}^-} {|x_{12}|_\mathfrak{q}^{\Delta_{12,3}}
   |x_{23}|_\mathfrak{q}^{\Delta_{23,1}} |x_{13}|_\mathfrak{q}^{\Delta_{31,2}}} \,,
}
where $|\cdot|_\mathfrak{q}$ now represents the $\mathfrak{q}$-adic norm.
Moreover, the constant coefficient $f_{123}^-$ exactly matches the regular tree answer~\cite{Gubser:2016guj,Gubser:2017tsi}.

As a final point of comparison, it is worth recalling the result in continuum AdS$_2$~\cite{Freedman:1998tz}, which takes an identical functional form,
\eqn{A3real}{
A_3^{\rm (AdS)}(x_1,x_2,x_3) = 
   \frac{1}{2}\frac{f_{123}^-} {|x_{12}|^{\Delta_{12,3}}
   |x_{23}|^{\Delta_{23,1}} |x_{13}|^{\Delta_{31,2}}
   }\,,
}
where the coefficient $f_{123}^-$ is obtained from a simple replacement in~\eqref{f123Def}: $\zeta^-_{\mathfrak{q}}(s) \rightarrow \zeta_{\infty}(s)$, where $\zeta_\infty(s)$ is the well-known local zeta function over $\mathbb{R}$, defined to be $\zeta_\infty(s) := \pi^{-s/2} \Gamma_{\rm Euler}(s/2)$.
\deleted{We will comment on the appearance of the local zeta functions~\eqref{LocalZeta} in Sec.~\ref{sec:DISCUSS}.}

\section{Bulk propagator identities}
\label{sec:PROPIDS}

In this section, we will work out some propagator identities that re-express bulk integrals over products of bulk-to-bulk propagators in terms of unintegrated combinations of bulk-to-bulk propagators. Such identities are useful in evaluating more complicated higher-point bulk diagrams involving exchanges~\cite{Jepsen:2019svc}.

\subsection{Two-propagator identity}
\label{sec:TWOPROP}

It is well-known in continuum AdS space (see, {\it e.g.}, Ref.~\cite{Hijano:2015zsa}) as well as the regular tree~\cite{Gubser:2017tsi} that a product of two bulk-to-bulk propagators meeting at a common point integrated over all of bulk space, can be written as a linear combination of bulk-to-bulk propagators without any integrals.
In this section, we will show that the same identity also continues to hold on the biregular tree.
More precisely,
\eqn{GGexp}{
\sum_{z \in T_{q_+,q_-}} \!\!\!\!\! G_{\Delta_1}(a,z) G_{\Delta_2}(z,b) = \frac{G_{\Delta_2}(a,b)-G_{\Delta_1}(a,b)}{m_{\Delta_1}^2-m_{\Delta_2}^2}.
}
This can be checked by explicitly evaluating the sum on the biregular tree geometry. The calculation is lengthy and not particularly illuminating, so we refrain from reproducing it here.
Instead, we will \replaced{provide an alternate ``derivation''}{establish this two-propagator identity} employing an approach that avoids doing bulk summations, which will efficiently extend to the three-propagator case in the following subsection.
\added{We say ``derivation'' in quotes because this approach may miss terms that are annihilated by derivative operators $(\square+m^2)$, but such contributions can be ruled out by performing all summations directly.}
 First, we start by assuming the following ansatz,
\eqn{GGansatz}{
\sum_{z \in T_{q_+,q_-}} \!\!\!\!\! G_{\Delta_1}(a,z) G_{\Delta_2}(z,b)  &= \alpha_1(q_a,q_b)\, G_{\Delta_1}(a,b) \cr 
&+ \alpha_2(q_a,q_b)\, G_{\Delta_2}(a,b)\,,
}
where the $\alpha_i(q_a,q_b)$s are undetermined parameters that are allowed to depend on the homogeneity degrees of vertices $a$ and $b$. The ansatz is inspired by the analogue of this identity that holds on the regular tree and continuum AdS spaces, where the $\alpha_i$s are constants.

Applying $(\square_{a}+m_{\Delta_1}^2)$ on the left-hand-side of~\eqref{GGansatz}, we get
\eqn{}{
\sum_{z \in T_{q_+,q_-}} \!\!\!\!\! \delta_{az} G_{\Delta_2}(z,b) = G_{\Delta_2}(a,b)\,,
}
while on the right-hand-side, we get
\eqn{}{
& (q_a+1+m_{\Delta_1}^2)\, \alpha_1(q_a,q_b) \,G_{\Delta_1}(a,b)  \cr 
-\,& \alpha_1(\tilde{q}_{a},q_b) \left(q_a  G_{\Delta_1}(\tilde{a}_+,b) 
+ G_{\Delta_1}(\tilde{a}_-,b) \right) \cr 
+\,& (q_a+1+m_{\Delta_1}^2) \, \alpha_2(q_a,q_b)\, G_{\Delta_2}(a,b)  \cr 
-\,& \alpha_2(\tilde{q}_{a},q_b) \left(q_a  G_{\Delta_2}(\tilde{a}_+,b) 
+ G_{\Delta_2}(\tilde{a}_-,b) \right).
}
where $\tilde{a}_{-}$ is the unique nearest-neighbour of vertex $a$ which is closer to $b$ while $\tilde{a}_{+}$ stands for any of the $q_{a}$ nearest-neighbours of $a$ which are further away from $b$ than $a$. Since $\alpha_1, \alpha_2$ only depend on the homogeneity degree of the nearest-neighbour vertices (which are all identical), they can be pulled outside as a common factor. 

In App.~\ref{APP:SPLIT}, we prove that if $c$ is any vertex on the geodesic joining $a$ and $b$, {\it i.e.} $c \in [a:b]$, then
\eqn{Gsplit}{
{G}_{\Delta}(a,b) = \frac{ {G}_{\Delta}(a,c) {G}_{\Delta}(c,b) }{ {G}_{\Delta}(c,c) }
\,.
}
Using this, we can re-express
\eqn{Gsplit1}{
 G_{\Delta}(\tilde{a}_{-},b) &= \frac{ G_{\Delta}(a, b) \, G_{\Delta}(\tilde{a}_{-},\tilde{a}_{-})}{G_{\Delta}(a,\tilde{a}_{-})} 
 = G_{\Delta}(a, b) \mathfrak{q}^{\Delta} \frac{   \psi_{\Delta}(q_{a})}{\psi_{\Delta}(\tilde{q}_{a}) }  
  \cr 
  &= G_{\Delta}(a, b)\, \mathfrak{q}^{2\Delta}\, \psi_{\Delta}(q_{a})^2   \,,
}
and
\eqn{Gsplit2}{
  G_{\Delta}(\tilde{a}_{+},b) &= \frac{ G_{\Delta}(a, b) \, G_{\Delta}(a,\tilde{a}_{+})}{ G_{\Delta}({a},{a})} 
= G_{\Delta}(a, b) \frac{1}{\mathfrak{q}^{\Delta}}  \frac{   \psi_{\Delta}(q_{a})}{\psi_{\Delta}(\tilde{q}_{a}) }
  \cr 
  &=   G_{\Delta}(a, b)\, \psi_{\Delta}(q_{a})^2 \,.
}
Substituting this back in the right-hand-side, we get
\eqn{}{
& (q_a+1+m_{\Delta_1}^2)\, \alpha_1(q_a,q_b)\, G_{\Delta_1}(a,b)  \cr 
-\,&  \left(q_a  + \mathfrak{q}^{2\Delta_1} \right)
\alpha_1(\tilde{q}_{a},q_b)\, \psi_{\Delta_1}(q_{a})^2\, G_{\Delta_1}(a,b) \cr 
+\,& (q_a+1+m_{\Delta_1}^2) \, \alpha_2(q_a,q_b)\, G_{\Delta_2}(a,b)  \cr 
-\,& \left(q_a  + \mathfrak{q}^{2\Delta_2} \right) \alpha_2(\tilde{q}_{a},q_b)\, \psi_{\Delta_2}(q_{a})^2\, G_{\Delta_2}(a,b).
}
Comparing with the left-hand-side, we require the coefficient of $G_{\Delta_1}(a,b)$ to vanish,
\eqn{}{
0 \stackrel{!}{=}   \alpha_1(q_a,q_b) 
-  \alpha_1(\tilde{q}_{a},q_b) \,,
}
where we used the definition of $\psi_\Delta$, which implies $\alpha_1$ does not depend on $q_a$.
Additionally, the coefficient of $G_{\Delta_2}(a,b)$ is required to be unity, yielding
\eqn{}{
 1 \stackrel{!}{=} 
 (q_a+1+m_{\Delta_1}^2) \alpha_2(q_a,q_b) \! -\! (q_a+1+m_{\Delta_2}^2) \alpha_2(\tilde{q}_{a},q_b) .
}
This equation holds for arbitrary $a$; thus, it also holds for its nearest neighbour $\tilde{a}$. This provides us with a second relation
\eqn{}{
 1 \stackrel{!}{=} 
 (\tilde{q}_a+1+m_{\Delta_1}^2) \alpha_2(\tilde{q}_a,q_b) \! -\! (\tilde{q}_a+1+m_{\Delta_2}^2) \alpha_2({q}_{a},q_b) .
}
Solving the set of linear equations for $\alpha_2$, we find
\eqn{alpha2}{
\alpha_2(q_a,q_b) &= \frac{{q}_a+\tilde{q}_a+2+m^2_{\Delta_1}+m^2_{\Delta_2}}{(\mathfrak{q}^{2\Delta_1}-\mathfrak{q}^{2\Delta_2}) (1-\mathfrak{q}^{2(1-\Delta_{12,})})} \cr 
 &= \frac{1}{m_{\Delta_1}^2-m_{\Delta_2}^2}\,,
}
where the second equality follows non-trivially from the mass-dimension relation. 
It is worthwhile restating this useful identity as follows,
\eqn{MassDiff}{
 \frac{q_+ + q_-+2 + m_{\Delta_1}^2 + m_{\Delta_2}^2}{(\mathfrak{q}^{2\Delta_1}-\mathfrak{q}^{2\Delta_2})(1-\mathfrak{q}^{2(1-\Delta_{12,})})} = \frac{1}{m_{\Delta_1}^2-m_{\Delta_2}^2} \,.
}
Finally, a symmetrical calculation yields,
\eqn{alphaOther}{
\alpha_1 = \frac{1}{m_{\Delta_2}^2-m_{\Delta_1}^2}\,.
}
Thus, we find the $\alpha_i$ coefficients are, in fact, $q_a,q_b$-independent constants that precisely reproduce~\eqref{GGexp}.

\subsection{Three-propagator identity}
\label{sec:THREEPROP}

In this section, we will present a generalisation of the three-point function of Sec.~\ref{sec:THREEPT} where all end-points are now contained in the bulk. Such a bulk summation appears as a subdiagram in several bulk Feynman diagrams.  

Specifically, consider the following sum on the biregular tree,\footnote{Note that we are considering normalised propagators in this section.}
\eqn{GGGdef}{
B_3(a_1,a_2,a_3):=\!\!\!\!\!\sum_{z \in T_{q_+,q_-}} \!\!\!\!\! G_{\Delta_1}(a_1,z)\, G_{\Delta_2}(a_2,z) \,   G_{\Delta_3}(a_3,z)\,,
}
where $a_1, a_2, a_3 \in T_{q_+,q_-}$ are three arbitrary bulk points. 
The three-point correlator~\eqref{A3Def} can be obtained from this in the limit where all $a_i$ approach the boundary.
On the biregular tree, we assume the following ansatz for the sum,\footnote{The ansatz~\eqref{GGGexp} is guided by the benefit of hindsight, motivated from the analogous identity on regular trees, the structure of the bulk summation in~\eqref{GGGdef} and the two-propagator identity derivation in the previous subsection. 

The product of propagators in the first term of~\eqref{GGGexp} can be rewritten without reference to the unique point of intersection $o$ and directly in terms of bulk-to-bulk propagators between bulk points $a_i$~\cite{Jepsen:2019svc}. It is straightforward to do so using the splitting identity~\eqref{Gsplit}, but we omit its inclusion for brevity.}
\eqn{GGGexp}{
B_3 =\;& \alpha_0(q_o)\, G_{\Delta_1}(a_1,o) G_{\Delta_2}(a_2,o) G_{\Delta_3}(a_3,o) \cr 
 &+ \alpha_1(q_{a_1})\, G_{\Delta_2}(a_1,a_2) G_{\Delta_3}(a_1,a_3) \cr 
& + \alpha_2(q_{a_2})\, G_{\Delta_1}(a_1,a_2) G_{\Delta_3}(a_2,a_3) \cr 
& + \alpha_3(q_{a_3})\, G_{\Delta_1}(a_1,a_3) G_{\Delta_2}(a_2,a_3) \,,
}
where $o={\rm join}(a_1,a_2,a_3)$ is the unique point of intersection of the geodesics joining $a_1,a_2,a_3$, and the coefficients $\alpha_i(q_{a_i})$ have a dependence on the bulk coordinates only via their homogeneity degree. We can determine these coefficients by explicitly performing the sum over the tree. However, this calculation is tedious and lengthy, so we refrain from including the details here. Instead, we will employ the alternate approach of the previous subsection to obtain three of the four coefficients without performing any bulk summations, assuming the form of the ansatz~\eqref{GGGexp}.

Acting with $(\square_{a_1}+m_{\Delta_1}^2)$ on~\eqref{GGGdef} and using the equation of motion for the bulk-to-bulk propagator, we get
\eqn{}{
(\square_{a_1}+m_{\Delta_1}^2) B_3 =  G_{\Delta_2}(a_1,a_2) G_{\Delta_3}(a_1,a_3)\,.
}
Applying the same operator on~\eqref{GGGexp}, we get
\eqn{}{
& (\square_{a_1}+m_{\Delta_1}^2) B_3 \cr 
&= (\square_{a_1}+m_{\Delta_1}^2) \! \left( \alpha_0(q_o)\, G_{\Delta_1}(a_1,o) G_{\Delta_2}(a_2,o) G_{\Delta_3}(a_3,o) \right) \cr 
&+ (\square_{a_1}+m_{\Delta_1}^2) \! \left( \alpha_1(q_{a_1}) G_{\Delta_2}(a_1,a_2) G_{\Delta_3}(a_1,a_3) \right) \cr 
&+ \alpha_2(q_{a_2}) \delta_{a_1a_2} G_{\Delta_3}(a_2,a_3) + \alpha_3(q_{a_3}) \delta_{a_1a_3} G_{\Delta_2}(a_2,a_3)\,.
}
Assuming $d(a_1,o)\geq 1$, the terms in the last line do not contribute. In this case, $o$ is also the point of intersection of geodesics joining $\tilde{a}_1, a_2, a_3$ for all nearest-neighbours $\tilde{a}_1$. Subsequently, the first term can be simplified to
\eqn{}{
 \alpha_0(q_o) \left( \!\!\right. & \left.  (\square_{a_1}+m_{\Delta_1}^2)  G_{\Delta_1}(a_1,o)  \right) G_{\Delta_2}(a_2,o) G_{\Delta_3}(a_3,o) \cr 
 &= \alpha_0(q_o) \delta_{a_1o} G_{\Delta_2}(a_2,o) G_{\Delta_3}(a_3,o) \,,
}
which does not contribute either, since we assumed $d(a_1,o)\geq 1$. Thus, we obtain the equation
\eqn{alpha1-eqn}{
& G_{\Delta_2}(a_1,a_2) G_{\Delta_3}(a_1,a_3) \cr 
\stackrel{!}{=}\;& (\square_{a_1}+m_{\Delta_1}^2) \left( \alpha_1(q_{a_1}) G_{\Delta_2}(a_1,a_2) G_{\Delta_3}(a_1,a_3) \right) ,
}
from which we will shortly extract $\alpha_1(q_{a_1})$. Applying the definition of the graph Laplacian~\eqref{Laplacian}, the right-hand-side of~\eqref{alpha1-eqn} becomes
\eqn{}{
& (q_{a_1}+1 + m_{\Delta_1}^2)  \alpha_1(q_{a_1})\, G_{\Delta_2}(a_1,a_2) G_{\Delta_3}(a_1,a_3)  \cr 
&- \alpha_1(\tilde{q}_{a_1}) \Big( G_{\Delta_2}(\tilde{a}_{1-},a_2) G_{\Delta_3}(\tilde{a}_{1-},a_3) \cr 
 &  +  q_{a_1}\,  G_{\Delta_2}(\tilde{a}_{1+},a_2) G_{\Delta_3}(\tilde{a}_{1+},a_3)  \Big),
}
where $\tilde{a}_{1-}$ is the unique nearest-neighbor of vertex $a_1$ which is closer to $a_2, a_3$ while $\tilde{a}_{1+}$ stands for any of the $q_{a_1}$ nearest-neighbors of $a_1$ which are further away from $a_2, a_3$ than $a_1$. Since $\alpha_1(\tilde{q}_{a_1})$ only depends on the homogeneity degree of the nearest-neighbour vertices (which are all identical), it can be pulled outside as a common factor.

Applying splitting identities such as~\eqref{Gsplit1}-\eqref{Gsplit2}, the right-hand-side of~\eqref{alpha1-eqn} simplifies to,
\eqn{}{
& G_{\Delta_2}(a_1,a_2)  G_{\Delta_3}(a_1,a_3) \cr 
\times & \Big[ (q_{a_1}+1 + m_{\Delta_1}^2)  \alpha_1(q_{a_1})   \cr 
& - \alpha_1(\tilde{q}_{a_1}) \psi_{\Delta_2}(q_{a_1})^2  \, \psi_{\Delta_3}(q_{a_1})^2  \left( q_{a_1} + \mathfrak{q}^{2\Delta_{23,}}   \right) \Big] .
}
Comparing with the left-hand-side of~\eqref{alpha1-eqn}, we require
\eqn{alpha1-eq1}{
 1 \stackrel{!}{=}  &\, (q_{a_1}+1 + m_{\Delta_1}^2)  \alpha_1(q_{a_1})  \cr 
&- \alpha_1(\tilde{q}_{a_1}) \psi_{\Delta_2}(q_{a_1})^2  \, \psi_{\Delta_3}(q_{a_1})^2  \left( q_{a_1} + \mathfrak{q}^{2\Delta_{23,}}  \right) .
}
The equation holds for arbitrary $a_1$ such that $d(a_1,o) \geq 1$. Thus it also holds for nearest-neighbors $\tilde{a}_1$ such that $d(\tilde{a}_1,o) \geq 1$. In that case, we get the constraint equation~\eqref{alpha1-eq1} with $q_{a_1} \leftrightarrow \tilde{q}_{a_1}$.
Solving the set of linear equations, we obtain 
\eqn{alpha1}{
\alpha_1(q_{a_1})  &= \frac{1}{m_{\Delta_1}^2-m_{\Delta_{23,}}^2} \cr 
&\times \frac{ \tilde{q}_{a_1}+1+m_{\Delta_1}^2 \!+ \displaystyle{\frac{\Theta_{23}(q_{a_1})}{\Theta_{23}(\tilde{q}_{a_1})}} \!\left(\! q_{a_1}+1 + m_{\Delta_{23,}}^2   \!\right) }{  \tilde{q}_{a_1} + 1 + m_{\Delta_1}^2 + q_{a_1} + 1+ m_{\Delta_{23,}}^2} ,
}
where we defined,
\eqn{}{
\Theta_{ij}(q_{a}) := \displaystyle{\frac{\psi_{\Delta_i}(q_{a})  \, \psi_{\Delta_j}(q_{a})}{\psi_{\Delta_{ij,}}(q_{a})} } \,,
}
so that the following identity holds,
\eqn{ThetaThetaT}{
\Theta_{ij}(q_a) \Theta_{ij}(\tilde{q}_a) = 1\,.
}
Using~\eqref{MassDiff}, one can also rewrite
\eqn{alpha1-again}{
\alpha_1(q_{a_1})  &= \frac{ \tilde{q}_{a_1}+1+m_{\Delta_1}^2 \!+ \displaystyle{\frac{\Theta_{23}(q_{a_1})}{\Theta_{23}(\tilde{q}_{a_1})}} \!\left(\! q_{a_1}+1 + m_{\Delta_{23,}}^2   \!\right) }{ (\mathfrak{q}^{2\Delta_1}-\mathfrak{q}^{2\Delta_{23,}})(1-\mathfrak{q}^{2(1-\Delta_{{123,}})}) }.
}
By symmetry
\eqn{alpha23}{
\alpha_2(q_{a_2}) &= \Big\{ \alpha_1(q_{a_1}) \quad {\rm \ with\ }(a_1 \leftrightarrow a_2, \Delta_1 \leftrightarrow \Delta_2) \Big\} \cr 
\alpha_3(q_{a_3}) &= \Big\{ \alpha_1(q_{a_1}) \quad {\rm \ with\ }(a_1 \leftrightarrow a_3, \Delta_1 \leftrightarrow \Delta_3) \Big\}\,.
    }

For the remaining coefficient, $\alpha_0(q_o)$, one does need to perform the bulk sum in~\eqref{GGGdef} and isolate terms proportional to $\prod_{i=1}^3 G_{\Delta_i}(a_i,o)$.
Fortunately, we have done this exact calculation already in Sec.~\ref{sec:THREEPT} in the context of the three-point correlator. The primary distinction compared to the calculation in Sec.~\ref{sec:THREEPT} is that the distances $d(a_i,o)$ are now finite. 
However, this does not affect the coefficient of $\prod_{i=1}^3 G_{\Delta_i}(a_i,o)$ in~\eqref{GGGexp}.\footnote{The remaining three terms in~\eqref{GGGexp}  we derived using the alternate method above arise precisely as a consequence of the finiteness of $d(a_i,o)$ and $d(a_i,a_j)$.
} The desired coefficient remains precisely the coefficient of $\prod_{i=1}^3 \hat{K}_{\Delta_i}(a_i,o)$ we computed earlier, given by 
\eqn{}{
\alpha_0(q_o) &= \left( f^-_{123} V^+_{123}(q_o) + f^+_{123} V^-_{123}(q_o) \right) \frac{\Psi_{123}(\tilde{q}_{o})}{\Psi_{123}(\tilde{q}_{v_0})} \cr 
&= (f_{123}^- W^+_{123} + (-1)^{\delta_{q_o,q_+}}  f_{123}^+ W^-_{123}) \Psi_{123}(\tilde{q}_{o})\,, 
}
where $V^\pm_{123}(q_o)$, $f^\pm_{123}$, and $W_{123}^\pm$ are defined in~\eqref{V123Def}, \eqref{f123Def}, and~\eqref{W123Def}, respectively.

In the regular tree limit, where $q_a=\mathfrak{q}$ for all vertices $a$, this result manifestly reduces to the result of Ref.~\cite{Gubser:2017tsi}, with {\it e.g.} $\alpha_1 = 1/(m_{\Delta_1}^2-m_{\Delta_{23,}}^2)$ and $\alpha_0 = f^-_{123}$. This follows immediately from the regular tree limit of the three-point function and the fact that in this limit, $\psi_\Delta(q_a) = 1/\mathfrak{q}^{\Delta/2}$ and $\Theta_{ij}(q_a) = 1$.

It is straightforward, though nontrivial, to check that in the coincident point limit ({\it e.g.} $a_1=a_2$), the three-propagator identity is consistent with the two-propagator identity~\eqref{GGexp}. The non-triviality of this check arises because the ratio
\eqn{Gratio}{ 
\frac{G_{\Delta_1}(a_1,z) G_{\Delta_2}(a_1,z)}{G_{\Delta_{12,}}(a_1,z)} = \frac{N_{\Delta_1} N_{\Delta_2}}{N_{\Delta_{12,}}} \frac{1}{\Theta_{12}({q}_{a_1}) \Theta_{12}(q_z)}
}
includes nontrivial dependence on the degree of bulk points $a_1$ and $z$. Consequently, for bulk points $z$ such that $d(a_1,z)= {\rm odd}$, the degree dependence on the right-hand-side drops out thanks to~\eqref{ThetaThetaT}, but no such simplification occurs for the remaining half of possibilities, when $d(a_1,z) = {\rm even}$. 
However, the identities remain non-trivially consistent.

Finally, we note that starting from the three-propagator identity~\eqref{GGGexp} and using~\eqref{GtoK}, it is straightforward to obtain the special cases where one or more of the bulk-to-bulk propagators are replaced by a bulk-to-boundary propagator. 
These should prove useful in evaluating higher-point or higher-loop bulk diagrams~\cite{Gubser:2017tsi,Jepsen:2019svc}.
 
\section{Discussion}
\label{sec:DISCUSS}

In this paper, we initiated the formulation of scalar field dynamics and holography on buildings (simplicial complexes) beyond regular trees or their quotients. 
Regular trees arise as symmetric spaces called Bruhat--Tits buildings associated with the isometry group ${\rm PGL}(2)$ over $p$-adic numbers.
Our primary focus was on biregular trees, which, as discussed in Sec.~\ref{sec:TREE}, are examples of Bruhat--Tits buildings associated with a $3\times 3$ unitary isometry group.
However, 
(bi)regular trees are part of a larger class of buildings associated with the group ${\rm PGL}(n)$ for all $n \geq 2$, which we expect to be natural playing grounds for holographic duality as well.

These ${\rm PGL}(n)$ buildings are also known as Euclidean buildings because certain maximal subspaces or `slices' of these buildings, called apartments,  correspond to tessellations of $\mathbb{R}^{n-1}$.
Thus, Euclidean buildings inherit certain features of flat space.
At the same time, they are naturally amenable to holography, having desirable properties such as a bulk depth direction that encodes the UV/IR bulk/boundary correspondence.
The nonarchimedean boundary of Euclidean buildings is described by the so-called Spherical buildings, a generalisation of the celestial sphere at infinity of Euclidean space~\cite{ronan2009lectures}.
It would be interesting to establish generalisations of our results to higher-dimensional buildings associated with ${\rm PGL}(n)$ for $n\geq 3$, as well as explore possible connections with flat space holography~\cite{Figueroa-OFarrill:2021sxz,Donnay:2023mrd,Pasterski:2023ikd,Bagchi:2025vri}. 
It would also be interesting to extend the results of this paper to spacetimes that are 
analogues of higher-dimensional black holes -- higher-genus objects obtained by quotienting higher-dimensional buildings by specific discrete subgroups -- along the lines of Refs.~\cite{Manin:2002hn,Heydeman:2016ldy,Heydeman:2018qty,Huang:2024ilq}.
There also exist a large class of higher-dimensional Hyperbolic buildings, whose apartments are tessellations of hyperbolic spaces instead, that were recently shown to admit holographic tensor networks~\cite{Gesteau:2022hss}.
A generalization of the present work to Hyperbolic buildings could also potentially provide new perspectives into differences between flat space and negatively curved space holography from a novel, discretised point of view.

 In this paper, we worked out the bulk-to-bulk and bulk-to-boundary propagators on the biregular tree, highlighting the similarities and novelties compared to the analogous results on regular trees. 
 Using these, we computed the two- and three-point correlators (as well as derived various bulk propagator identities)
  for the putative boundary dual of an effective scalar field theory in the bulk, where the conformal group is replaced by a $3 \times 3$ unitary group over a quadratic extension of $p$-adic numbers. 
 The three-point function, presented in~\eqref{ThreePt}, displays several expected and novel features. 
 It has the expected scaling of a three-point correlator of a standard CFT.
 However, due to the distinct semihomogeneous nature of biregular tree bulk geometry, a new three-point ``tensor structure'' emerges, that depends on the homogeneity degree of the unique bulk point of intersection of geodesics joining the three boundary insertion points.
 Such a structure is absent in regular tree or continuum geometries. 
 In the limit where the biregular tree collapses to the regular tree, the tensor structure disappears, and we reproduce the expected three-point conformal structure of a ($p$-adic) conformal field theory.

Moreover, we obtained closed-form expressions for the three-point coefficients at tree-level.
When expressed in terms of appropriate local zeta functions, these coefficients take a form very similar to analogous results on the regular tree and in continuum space.
However, on the biregular tree, a notable feature in the three-point coefficients is the appearance of two types of local zeta functions, $\zeta^\pm_{\mathfrak{q}}$. 
It turns out the same coefficients can also be expressed in terms of local zeta functions $\zeta^-_{\mathfrak{q}^2}$, {\it i.e.}, local zeta functions with prime $\mathfrak{q}$ replaced by $\mathfrak{q}^2$.
Intriguingly, these local zeta functions also arise in the study of zeta functions associated with the trivial and nontrivial multiplicative sign characters in the unramified quadratic extension of the $\mathfrak{q}$-adic field~\cite{Gelfand:1968, Brekke:1993gf, Gubser:2017qed,Gubser:2018cha}. 

This connection is particularly suggestive given that the algebraic Bruhat--Tits construction of the biregular tree involves precisely the unramified quadratic extension (as described in Sec.~\ref{sec:PADIC}). 
A detailed study into the algebraic origins of these results may yield valuable insights, potentially providing new perspectives on continuum holography as well.

A natural next step is to compute four-point correlators and higher, along the lines of Refs.~\cite{Gubser:2017tsi,Jepsen:2018dqp,Jepsen:2018ajn,Jepsen:2019svc} to furnish further insights into the holographic duality. 
Propagator identitites of Sec.~\ref{sec:PROPIDS} would prove useful in this task.
The structure of the four-point contact and exchange diagrams will offer a window into the OPE structure of the putative dual theory on the boundary, especially how the tensor structures of the unitary symmetry group organise to produce an analogue of conformal block decomposition.

From the purely boundary perspective, a crucial task is to formulate ``conformal'' field theory on the boundary of the biregular tree, an ultrametric fractal-like space, along the lines of Ref.~\cite{Melzer1989}, employing recent advances in the study of derivative operators on such spaces~\cite{bradley2025}.
An alternative approach to deriving the boundary action by integrating out the bulk~\cite{Zabrodin1989} is also worth investigating.
In either case, it would be useful to develop Fourier methods on the fractal boundary, which would have applications in momentum space analysis of quantum field theories on the boundary as well as in the precise computation of the normalisation of the two-point function, as mentioned in Sec.~\ref{sec:TWOPT}.

It would also be interesting to study nonarchimedean string theory~\cite{Freund:1987kt,Freund:1987ck,Zabrodin1989} wherein the worldsheet is described by a biregular tree.
Given that (bi)regular trees could be viewed as subsets of the Berkovich projective line~\cite{baker2008introduction}, a $p$-adic analytic space, this suggests yet another avenue for generalisation towards developing arithmetic analogues of holography and string theory~\cite{Goldfarb:2023vpx,heckman2023speculations}.

Another possible application of our work involves hyperbolic fracton models on tessellations of hyperbolic disks~\cite{Yan:2018nco,Yan:2019quy} by alternate tilings of two distinct regular polygons, such as the hexaoctogonal tessellation. The subsystem symmetries of such tessellations are described by biregular trees, allowing the possibility of a dual description for the low-temperature limit in terms of infinitely many decoupled copies of a massless scalar field on biregular trees~\cite{Yan:2023lmj}.
It would be interesting to establish this duality and study its consequences.

We hope to address these questions in the future.

\subsection*{Acknowledgments}

 The work of SP was supported in part by SERB (now ANRF) grant SRG/2023/001475, Department of Science and Technology, Government of India.
The work of AM was supported by the SRF fellowship of the Council of Scientific and Industrial Research (CSIR), India.

\appendix

\section{Splitting identities}
\label{APP:SPLIT}

In this appendix, we derive the splitting identities \eqref{Ksplit} and \eqref{Gsplit} used in the main text.

\subsection{Bulk-to-boundary splitting}

Let $z$ be an arbitrary bulk vertex and $x$ a boundary point. Let $w$ be another bulk point such that $w \in (x:z]$. 
Then, it is easily checked that,
\eqn{}{
\langle z,x \rangle_{v_0} = \langle w,x \rangle_{v_0} - d\left(z,w\right).
}
Therefore, the bulk-to-boundary propagator~\eqref{Khat} between $z$ and $x$ takes the form,
\eqn{}{
\hat{K}_{\Delta}(z,x) = \frac{\psi_{\Delta}(\tilde{q}_z)}{\psi_{\Delta}(\tilde{q}_{v_0})}\, \mathfrak{q}^{\Delta \langle w,x\rangle_{v_0}}\,\mathfrak{q}^{-\Delta d(z,w)}\,.
}
The right-hand-side can be rearranged suggestively as follows,
\eqn{}{
 \left(\frac{\psi_{\Delta}(\tilde{q}_w)}{\psi_{\Delta}(\tilde{q}_{v_0})}\, \mathfrak{q}^{\Delta \langle w,x\rangle_{v_0}}\!\right)\! \left( \!\frac{\psi_{\Delta}(\tilde{q}_z)}{\psi_{\Delta}(q_w)}\,\mathfrak{q}^{-\Delta d(z,w)}\!\right) 
 \frac{\psi_{\Delta}(q_w)}{\psi_{\Delta}(\tilde{q}_w)}\,,
}
to furnish the splitting identity,
\eqn{Ksplit1}{
\hat{K}_{\Delta}(z,x) = \frac{\hat{K}_{\Delta}(w,x)\, \hat{G}_{\Delta}(w,z)}{\hat{G}_{\Delta}(w,w)}\,,
}
where $\hat{G}_{\Delta}$ is the unnormalized bulk-to-bulk propagator~\eqref{Ghat}.
For bulk points $w$ such that $z \in (x:w]$, the corresponding splitting identity is obtained from~\eqref{Ksplit1} by switching $z \leftrightarrow w$ and rearranging,
\eqn{Ksplit2}{
\hat{K}_{\Delta}(z,x) = \frac{\hat{K}_{\Delta}(w,x)\, \hat{G}_{\Delta}(z,z)}{\hat{G}_{\Delta}(w,z)}\,,
}
which completes the derivation of~\eqref{Ksplit}.

\subsection{Bulk-to-bulk splitting}

Let $c \in [a:b]$ be any vertex on the geodesic joining bulk vertices $a$ and $b$. 
Then,
\eqn{dist}{
d\left(a,b\right) = d\left(a,c\right) + d\left(c,b\right)\,,
} 
so that the (normalised) bulk-to-bulk propagator~\eqref{G} becomes
\eqn{}{
G_{\Delta}(a,b) &= N_{\Delta} \frac{\psi_\Delta(\tilde{q}_a) }{ \psi_\Delta(q_b) }  \, \mathfrak{q}^{-\Delta\, \left(d(a,c) + d(c,b)\right)} \cr 
&= \frac{\left(\! N_{\Delta} \frac{\psi_\Delta(\tilde{q}_a) }{ \psi_\Delta(q_c) }  \, \mathfrak{q}^{-\Delta\, d(c,b)} \! \right) \!\! \left( \! N_{\Delta} \frac{\psi_\Delta(\tilde{q}_c) }{ \psi_\Delta(q_b) }  \, \mathfrak{q}^{-\Delta\, d(a,c)} \! \right)}{N_{\Delta}\displaystyle{\frac{\psi_\Delta(\tilde{q}_c) }{ \psi_\Delta(q_c) }} },
}
 which we recognise to be the form of the splitting identity~\eqref{Gsplit}.

\vspace{2em}
\bibliographystyle{ssg}
\bibliography{main} 
\end{document}